# APPROXIMATING (UNWEIGHTED) TREE AUGMENTATION VIA LIFT-AND-PROJECT, PART II

JOSEPH CHERIYAN AND ZHIHAN GAO

ABSTRACT. In Part II, we study the unweighted *Tree Augmentation Problem* (TAP) via the Lasserre (Sum of Squares) system. We prove that the integrality ratio of an SDP relaxation (the Lasserre tightening of an LP relaxation) is $\leq \frac{3}{2} + \epsilon$, where $\epsilon > 0$ can be any small constant. We obtain this result by designing a polynomial-time algorithm for TAP that achieves an approximation guarantee of $(\frac{3}{2} + \epsilon)$ relative to the SDP relaxation. The algorithm is combinatorial and does not solve the SDP relaxation, but our analysis relies on the SDP relaxation. We generalize the combinatorial analysis of integral solutions from the previous literature to fractional solutions by identifying some properties of fractional solutions of the Lasserre system via the decomposition result of Karlin, Mathieu and Nguyen (IPCO 2011).

## 1. INTRODUCTION

In the weighted *Tree Augmentation* problem we are given a connected, undirected multigraph $G$ with non-negative costs on the edges, together with a spanning tree $T = (V(G), \widehat{E}_T)$ of $G$; the goal is to find a set of edges, $F \subseteq E(G) - \widehat{E}_T$, of minimum cost such that $(V, \widehat{E}_T \cup F)$ is 2-edge connected. By a link we mean an element of $E(G) - \widehat{E}_T$; thus, a link is an edge of $G$ that can be used to augment $T$. We say that a link $uw$ covers an edge $\hat{e} \in \widehat{E}_T$ (a tree-edge) if $T + uw - \hat{e} = (V, \widehat{E}_T \cup \{uw\} - \{\hat{e}\})$ is connected. We say that a set of links $F$ covers the tree $T$ if every edge of $T$ is covered by at least one link of $F$; it can be seen that $F$ covers $T$ iff $(V, \widehat{E}_T \cup F)$ is 2-edge-connected. Thus, the goal is to compute a set of links of minimum cost that covers $T$.

The weighted Tree Augmentation problem was first studied by Frederickson and Jaja in 1981 [4]. They showed that the problem is NP-hard, and they presented a 2-approximation algorithm. Subsequently, it has been proved that even the unweighted Tree Augmentation problem is APX-hard, see [9, Section 4]; thus the problem has no PTAS assuming P≠NP.

There have been some important advances on this problem for the corresponding *unweighted* (i.e., uniform weight) problems. Following [5], we use the abbreviation TAP for the *unweighted* Tree Augmentation problem. Nagamochi, see [6], presented the first algorithm for TAP that improved on the approximation guarantee of 2; the approximation guarantee is ≈ 1.875. Subsequently, Even, et al., [5] built on the ideas and techniques initiated by Nagamochi and presented an elegant algorithm and analysis that achieves an approximation guarantee of 1.8. In a conference publication from 2001, Even, et al., reported a 1.5 approximation algorithm for TAP, see the extended abstract [3], and recently, Kortsarz & Nutov [10] presented the journal version of this result.

We prove that the integrality ratio of an SDP relaxation (the Lasserre tightening of an LP relaxation) is $\leq \frac{3}{2} + \epsilon$, where $\epsilon > 0$ can be any small constant. We obtain this result by designing a polynomial-time algorithm for TAP that achieves an approximation guarantee of $(\frac{3}{2} + \epsilon)$ relative to the SDP relaxation. The algorithm is combinatorial and does *not* solve the SDP relaxation, but our analysis relies on the SDP relaxation. Observe that our integrality ratio (and approximation guarantee) is proved relative to a "weaker lower bound" than the optimum value (since the feasible region of a relaxation is a superset of the convex full of integer solutions). Moreover, the approximation guarantee of 1.5 follows as a corollary of our main result (but, we cannot eliminate the "$+\epsilon$" in our bound on the integrality ratio by this corollary).

Linear programming relaxations for the weighted version of TAP have been studied for many years. (We use the standard abbreviation LP to mean a linear programming relaxation or a linear programming problem.) There is an obvious "covering" LP: we have a variable $x_e$ for each edge $e \in E(G) - \widehat{E}_T$ and we have a covering constraint for each edge of $T$. It is well known that the integrality ratio of this LP is $\leq 2$; this can be deduced from Jain's result [7]. A lower bound of 1.5 on the integrality ratio is known [1]; in fact, the construction for the lower bound uses uniform weights for the edges in $E(G) - \widehat{E}_T$, hence, the lower bound applies for TAP. We formulate an LP that is a tightening of the obvious "covering" LP for TAP; see ($LP_0$) in Section 3.

The Lasserre system applies to an initial LP, and it derives a sequence of tightenings of the initial LP; these tightened relaxations are indexed by a number $t = 0, 1, \dots$ called the *level*, where the level 0 tightening means the initial LP. A key "decomposition theorem" (see Theorem 4.1, [11, 8]) asserts that a feasible solution at level $t$ can be written as a convex combination of feasible solutions at a lower level such that all of these lower-level solutions $y$ are "locally integral." Here, "locally integral" means that there is a specified subset $J \subseteq E(G) - \widehat{E}_T$, such that the solution $y$ takes only zero or one values on this subset (i.e., $y_e \in \{0, 1\}, \forall e \in J$). A key point is that the difference in levels (between the level $t$ of the given feasible solution and the level of the lower-level "locally integral" solutions) does *not* depend on the size of $J$, rather, it depends on the following "combinatorial parameter" determined by $J$. Suppose that there exists a constant $k$ such that every feasible solution $x$ of the initial LP has $\leq k$ entries in $J$ that have value one, i.e., $|\{e \in J : x_e = 1\}| \leq k$ for every feasible solution $x$ of the initial LP. Then for any $t > k$, a feasible solution at level $t$ can be written as a convex combination of "locally integral" feasible solutions at level $(t - k)$. This property does not hold for other weaker Lift-and-Project systems such as the Lovász-Schrijver system or the Sherali-Adams system.

The analysis of our algorithm is based on a potential function. Our potential function is derived from the Lasserre tightening of the initial LP.

Our algorithm is "combinatorial" and we do *not* need to solve the initial LP nor its Lasserre tightening to run the algorithm; but, the analysis of the algorithm relies on the Lasserre tightening. Our algorithm essentially follows the "algorithmic scheme" of [5, Section 3.4], see Section 7 for details. The algorithm is a greedy-type iterative algorithm that makes a leaves-to-root scan over the tree $T$ and (incrementally) constructs a set of links $F$ that covers $T$. The algorithm starts with $F := \emptyset$, at each major step it adds one or more links to $F$ (it never removes links from $F$), and at termination, it outputs a set of links $F$ that covers $T$ such that $|F| \leq$ the potential function. The algorithm incurs a cost of one unit for each link added to $F$. The key to the analysis is to show that for each major step, the cost incurred (i.e., one plus the number of links added to $F$) is compensated by a part of the potential function.

Informally speaking, our analysis in Section 8 asserts the following:

> if the naive algorithm gets "stuck" then the instance contains a small combinatorial obstruction, a so-called deficient tree, see Theorems 8.7, 8.8.

This assertion is the key to this paper; it turns out that the algorithmic aspects as well as the analysis of the approximation guarantee are straightforward consequences. Our analysis in Section 8 makes essential use of the Lasserre system and the decomposition theorem, see Figure 1.

For ease of exposition and for the sake of motivation, this paper has several discussions that contain forward references. But, these forward references are not relevant for the correctness of our arguments; indeed, all discussions containing forward references could be deleted, and this will have no effect on the validity of our proofs. Also, Part II repeats some definitions, figures, and discussions from Part I, but proofs from Part I are not repeated.

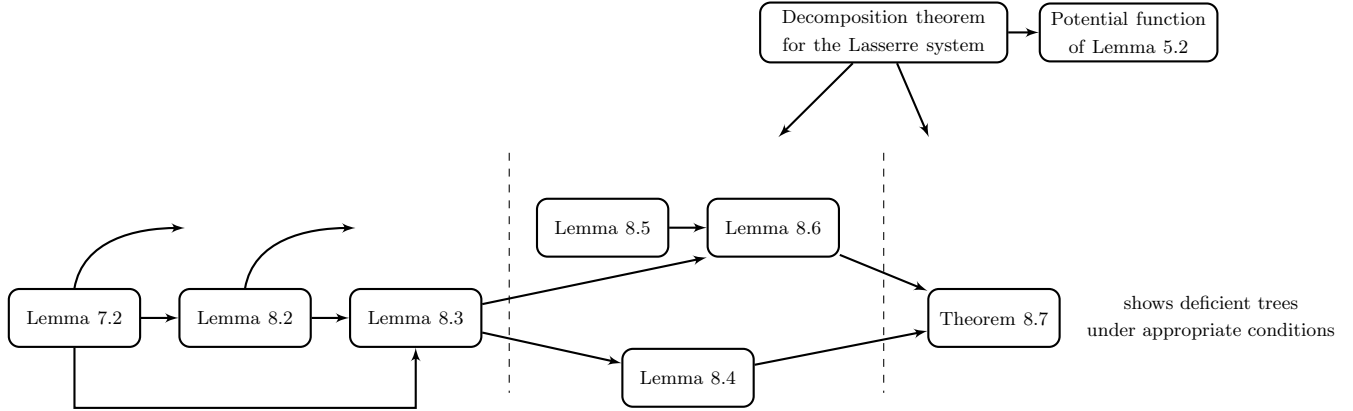

FIGURE 1. Illustration of the role of the decomposition theorem for the Lasserre system (Theorem 4.1). Our analysis consists of three blocks of results: low-level, middle-level, and high-level; these are shown from left to right in the figure. Note that Lemmas 7.2, 8.2 are often used in the proofs of subsequent results.

An outline of Part II is as follows. Section 2 has definitions and notation; several items are new to Part II (and not needed in Part I). We adopt the notation and terms of Even, et al., [5], where possible; this will aid readers familiar with that paper. Section 3 presents the initial LP; this is the same LP as in Part I. Section 4 discusses the Lasserre tightening of the initial LP, and proves some basic properties and inequalities; most of these results are from Part I, but there are two new results. Section 5 derives our potential function, based on a solution $y$ of the Lasserre tightening; this potential function differs significantly from the potential function of Part I. Section 6 starts the presentation of the algorithm (and credits) by elaborating on two preprocessing steps. Section 7 completes the discussion of the algorithm (and credits) by presenting the main loop of the algorithm. The most important component of this paper is Section 8; this section presents the analysis of the algorithm by first proving some low-level properties, then builds on this to prove some intermediate-level lemmas, and then proves the key theorem on deficient trees (Theorem 8.7); this section also presents and proves the last piece of the algorithm, namely, the handling of deficient trees.

2. PRELIMINARIES AND NOTATION

This section presents definitions and notation.

**Standard notation including tree $T$, link set $E$.** Let $G = (V, E(G))$ be a connected, undirected multigraph, and let $T = (V, \widehat{E}_T)$ be a spanning tree of $G$. We assume that $|V| \geq 2$. By a *tree-edge* we mean an edge of $T$. Let $E$ denote the edge-set $E(G) - \widehat{E}_T$; we call $E$ the *link set* and we call an element $\ell \in E$ a *link*; thus, a link is an edge of $G$ that can be used to augment $T$. An instance of TAP consists of $G$ and $T$. We assume that all instances of interest have feasible solutions, that is, we assume that $(V, \widehat{E}_T \cup E)$ is 2-edge connected. The goal is to find a minimum-size subset $F$ of $E$ such that augmenting $T$ by $F$ results in a 2-edge connected multigraph, i.e., $(V, \widehat{E}_T \cup F)$ is 2-edge connected.

For two nodes $u, v \in V$, we use $P_{u,v} = P_{v,u}$ to denote the unique path of the tree $T$ between $u$ and $v$.

For a node $v \in V$, we denote the number of tree-edges incident to $v$ by $\deg_T(v)$. For any node $v \in V$, we use $\delta_E(v)$ to denote the set of links incident to $v$.

For any $U \subseteq V$, we denote the set of links with both ends in $U$ by $E(U)$, and for any two subsets $U, W$ of $V$, we denote the set of links with one end in $U$ and the other end in $W$ by $E(U, W)$; thus, $E(U, W) := \{uw \in E : u \in U, w \in W\}$. We use similar notation for some subsets of $E$; for example, $E^{reg}$ denotes a particular subset of $E$ (defined below), and $E^{reg}(U, W)$ denotes $\{uw \in E^{reg} : u \in U, w \in W\}$.

We say that a link $uw$ *covers* a tree-edge $\hat{e}$ if $P_{u,w} \ni \hat{e}$. Similarly, we say that a subtree of $T$ is *covered* by a set of links $J \subseteq E$ if each tree-edge of the subtree is covered by some link of $J$. For any tree-edge $\hat{e} \in \widehat{E}_T$, we use $\delta_E(\hat{e})$ to denote the set of links that cover $\hat{e}$, thus, $\delta_E(\hat{e}) = \{uw \in E : \hat{e} \in P_{u,w}\}$.

**Shadows and the shadow-closed property.** For two links $u_1 v_1$ and $u_2 v_2$, if $P_{u_1, v_1} \subseteq P_{u_2, v_2}$, then we call $u_1 v_1$ a *shadow* (or, *sublink*) of $u_2 v_2$.

For any link $uv \in E$, if all sublinks of $uv$ also exist in $E$, then we call $E$ *shadow-closed*. Clearly, if $E$ is not shadow-closed, then we can make it shadow-closed by adding all sublinks of each of the original links. It can be seen that this preserves the optimal value of any instance of TAP.

Following Even, at al., [5], we make the next assumption (see Assumption 2.2 of [5]).
**Assumption:** $E$ is shadow-closed.

**Root, ancestor, descendant and rooted subtrees.** One of the nodes $r$ of $T$ is designated as the root; thus, we have a rooted tree $(T, r)$.

Let $v$ be a node of $T$. If a node $w$ belongs to the path $P_{v,r}$, then $w$ is called an *ancestor* of $v$, and $v$ is called a *descendant* of $w$. If a descendant $w$ of $v$ is adjacent to $v$ (thus, $w \neq v$), then $w$ is called a *child* of $v$, and $v$ is called a *parent* of $w$. Clearly, every node in $V - \{r\}$ has a unique parent. If a node $v$ has no child, then we call $v$ a *leaf*; if $v$ has no child, then $\deg_T(v) = 1$. Note that $r$ has at least one child (since $|V| \geq 2$), thus, $r$ is not a leaf, even if $\deg_T(r) = 1$. Throughout, we use $L$ to denote the set of leaves; thus, $L = \{v \in V : v \text{ is a leaf of } T\}$.

For any leaf $v$ of $T$, $up(v)$ denotes a node $q$ in $P_{v,r}$ that is nearest to the root and adjacent to $v$ via a link. Since $E$ is shadow-closed, it can be seen that $up(v)$ is an ancestor of $v$, for each leaf $v$.

For any node $v$, we use $T_v$ to denote the rooted subtree of $(T, r)$ induced by $v$ and its descendants. We use $L(T_v)$ to denote the set of leaves of the subtree $T_v$. Throughout, the terms *tree* or *subtree* refer to a rooted subtree of $(T, r)$ that is "closed w.r.t. descendants" (if a node $w$ is in the subtree, then all descendants of $w$ are in the subtree).

**Property 2.1.** *Suppose that $\bar{T}$ is a rooted tree. Let $\bar{T}_{v_1}$ and $\bar{T}_{v_2}$ be two (rooted) subtrees of $\bar{T}$. Then $\bar{T}_{v_1}$ and $\bar{T}_{v_2}$ either share no node or one is contained in the other.*

**Remark 2.2.** *Throughout, symbols such as $\bar{T}$, $\bar{J}$, etc. denote an arbitrary item/set (rather than the complement of another set).*

**Stems and twin links.** We call a node $s$ of $T$ a *stem* if $s \neq r$, $s$ has exactly two children, $s$ has exactly two descendants that are leaves, and there exists a link in $E$ between the two leaves of $T_s$; we call the link between the two leaves of $T_s$ a *twin link*, and denote it by $twinlk(s)$. (Our definition of stem differs slightly from [5, Definition 3.1].) Let $E^{twin}$ denote the set of twin links. Observe that there is a one-to-one correspondence between twin links and stems. Throughout, we use $\mathcal{S}$ to denote the set of stems; thus, $\mathcal{S} = \{v \in V : v \text{ is a stem of } T\}$. Moreover, we use $\mathcal{R}$ to denote the set of nodes that are neither stems nor leaves; thus $\mathcal{R} = V - (\mathcal{S} \cup L)$. (The notion of stems and twin links is due to [6].)

For any stem node $s$, we define $\delta_E^{\text{UP}}(s) = \{vs \in E : v \notin V(T_s)\}$. Similarly, we define $\delta_E^{\text{DN}}(s) = \{vs \in E : v \in V(T_s)\}$.

**Contraction and compound nodes.** We use the standard notion of *contracting* a link or a set of links, see [5], [2, Chapter 1]. We use $T' := T/F$ to denote the tree obtained by contracting each of the 2-edge connected components of $T + F = (V, \widehat{E}_T \cup F)$ to a single node. Each of the contracted nodes of $T' = T/F$ is called a *compound node*, see [5, Section 3.2]; thus, each compound node corresponds to a set of two or more nodes of $V(T)$. Each of the other nodes of $T/F$ is called an *original node*.

We also call a node of $T$ an *original node*. We call a link in $E$ an *original link*. When we say that $\ell$ is a *link w.r.t. $T$* we mean that $\ell$ is an original link; when we say that $\ell'$ is a *link w.r.t. $T' = T/F$*, we mean that there exists an original link $\ell$ whose image in $T/F$ is $\ell'$.

**Buds and buddy links.** Besides stems and their associated subtrees, one other type of subtree (and some of the incident links) plays an important role in this paper.

We call a leaf $b_0$ a *bud* (see Figure 2) if there exists a (rooted) subtree $T_v$ with exactly three leaves $b_0, b_1, b_2$ such that (i) $up(b_0)$ is a descendant of $v$ (possibly, $up(b_0) = v$), (ii) $T_{up(b_0)}$ contains a stem $s$ such that $b_0, b_1$ are the leaves of $T_s$, and (iii) the link $b_1 b_2$ exists. We call $b_1 b_2$ the *buddy link* of $b_0$ and denote it by $buddylk(b_0)$. Observe that there exists an ancestor $q$ of $s$ in $T_v$ such that $q$ is the least common ancestor of $s$ and $b_2$; possibly, $q = v$, and possibly, $q = v = r$; moreover, $L(T_{up(b_0)}) = \{b_0, b_1\}$ or $L(T_{up(b_0)}) = \{b_0, b_1, b_2\}$.

Note that $b_1$ may be a bud as well. In that case, each leaf of $T_{up(b_1)}$ is in $\{b_0, b_1, b_2\}$, there exists a link between $b_0, b_2$, and $buddylk(b_1) = b_0 b_2$.

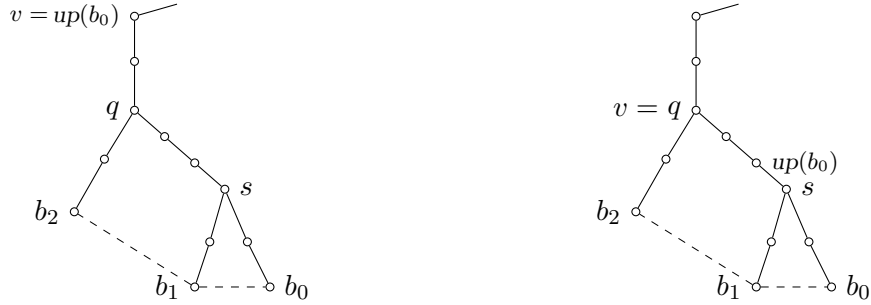

FIGURE 2. Illustration of a bud $b_0$ and the subtree $T_v$. The figure on the left shows the case when $up(b_0)$ is an ancestor of all three leaves in $T_v$. The figure on the right shows the other case.

Consider a bud $b_0$, and let $s, q$ be as above; we define
$$\mathcal{R}^{special}(b_0) := V(P_{b_0, up(b_0)}) - V(P_{s,q}) - \{b_0\}.$$

Thus $\mathcal{R}^{special}(b_0)$ contains the internal nodes in the tree-path between $b_0$ and $s$; moreover, if $up(b_0)$ is an ancestor of all three leaves $b_0, b_1, b_2$, then $\mathcal{R}^{special}(b_0)$ also contains all nodes on the tree-path between the parent of $q$ and $up(b_0)$; there are no other nodes in $\mathcal{R}^{special}(b_0)$ (see Figure 3).

**Fact 2.3.** $\mathcal{R}^{special}(b_0)$ *is the set of nodes $w$ such that there exists a link $b_0 w$ in $E$, and $w$ is not on the tree-path between the two ends of $buddylk(b_0)$. Every node in $\mathcal{R}^{special}(b_0)$ has a unique child, and is an ancestor of $b_0$. For two buds $b, \bar{b}$ that are not the leaves of the same subtree $T_s$, where $s$ is a stem, the sets $\mathcal{R}^{special}(b)$ and $\mathcal{R}^{special}(\bar{b})$ are disjoint. (Thus, $\mathcal{R}^{special}(b) \cap \mathcal{R}^{special}(\bar{b}) = \emptyset$, unless the least common ancestor (in $T$) of $b, \bar{b}$ is a stem $s$.)*

We denote the set of buds and the set of buddy links by $L^{bud}$ and $E^{buddy}$, respectively. Observe that there is a unique buddy link for each bud; thus, there is a bijection between $L^{bud}$ and $E^{buddy}$.

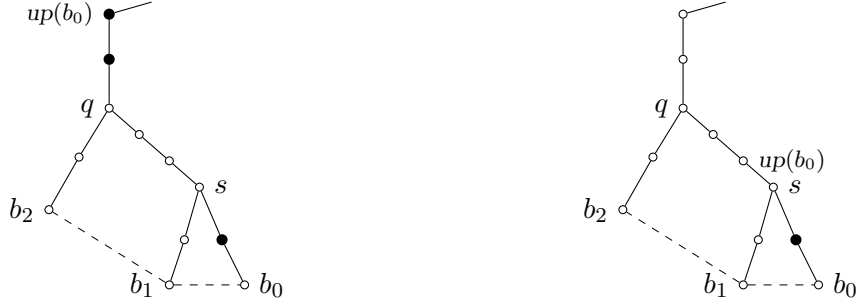

FIGURE 3. Illustration of $\mathcal{R}^{special}(b_0)$ for a bud $b_0$. The nodes in $\mathcal{R}^{special}(b_0)$ are indicated by solid circles. The figure on the left shows the case when $T_{up(b_0)}$ has three leaves. The figure on the right shows the other case.

For any node $w$ of the tree $T$, we denote by $L^{bud}(w)$ the set of buds in the subtree $T_w$. For a stem $s$, note that $L^{bud}(s)$ may contain zero, one, or two nodes.

Let $\mathcal{R}^{special} = \cup_{b \in L^{bud}} \mathcal{R}^{special}(b)$ and $\mathcal{R}^{nonspcl} = \mathcal{R} - \mathcal{R}^{special}$. Thus, we partition the set $\mathcal{R}$ (of nodes that are neither stems nor leafs) into two subsets, the "special" subset $\mathcal{R}^{special}$ and the "normal" subset $\mathcal{R}^{nonspcl}$.

We use $E^{reg}$ to denote the set $E - (E^{twin} \cup E^{buddy})$, namely, the set of links that are neither twin links nor buddy links.

**Vectors and convex combinations.** For any vector $x \in \mathbb{R}^E$, let $ones(x)$ denote the set of links of $x$-value one, thus $ones(x) = \{uv \in E : x_{uv} = 1\}$.

For a vector $x \in \mathbb{R}^E$ and any subset $J$ of $E$, $x(J)$ denotes $\sum_{e \in J} x_e$. Given several vectors $v^1, v^2, \ldots$, we write one of their convex combinations as $\sum_{i \in Z} \lambda_i v^i$; thus, $Z$ is a set of indices, and we have $\lambda_i \geq 0, \forall i \in Z$, and $\sum_{i \in Z} \lambda_i = 1$.

## 3. INITIAL LP RELAXATION

This section presents our initial LP relaxation of shadow-closed instances of TAP, denoted by $(LP_0)$; this is the same LP as in Part I (Section 3).

Let $u_1v_1$ and $u_2v_2$ be a pair of links. We call it an *overlapping pair* of links if (i) $P_{u_1,v_1}, P_{u_2,v_2}$ have one or more tree-edges in common, and (ii) either an end of $u_1v_1$ is in $P_{u_2,v_2}$, or an end of $u_2v_2$ is in $P_{u_1,v_1}$. We call a set of links $J$ an *overlapping clique* if every pair of links in $J$ is an overlapping pair. We use the notion of overlapping pairs for stating $(LP_0)$.

$$
\begin{aligned}
(\boldsymbol{LP}_0) \quad \text{minimize} : & \sum_{uv \in E} x_{uv} \\
\text{subject to} : & \sum_{uv \in \delta_E(\hat{e})} x_{uv} \geq 1, \quad \forall \hat{e} \in \widehat{E}_T \\
& x_{u_1v_1} + x_{u_2v_2} \leq 1, \quad \forall \text{ overlapping pairs } \ u_1v_1, u_2v_2 \in E \\
& 0 \leq x \leq 1
\end{aligned}
$$

## 4. LASSERRE TIGHTENING AND ITS PROPERTIES

This section presents the Lasserre tightening of the initial LP and proves some basic properties and inequalities; most of these results are from Part I, but there are two new results. The proofs of the results from Part I are omitted.

Let $\text{LAS}_t(LP_0)$ denote the level $t$ tightening of $(LP_0)$ by the Lasserre system.[1] Recall that $ones(x)$ denotes the set of links of $x$-value one, where $x \in \mathbb{R}^E$ is a feasible solution of $(LP_0)$. Rothvoß, see [11, Theorem 2], formulated the following decomposition theorem for the Lasserre system, based on an earlier decomposition theorem due to Karlin-Mathieu-Nguyen [8]. (We use this particular formulation and not the original statement of [8]; hence, we reference both [8] and [11].)

**Theorem 4.1.** *Let $J \subseteq E$. Let $k$ be a positive integer such that $|ones(x) \cap J| \leq k$ for every feasible solution $x$ of $(LP_0)$. Then for every feasible solution $y \in \text{LAS}_t(LP_0)$, where $t \geq k+1$, $y$ can be written as a convex combination $y = \sum_{i \in Z} \lambda_i x^i$ such that $x^i$ is in $\text{LAS}_{t-k}(LP_0)$ and $x^i|_J$ is integral (i.e., $x^i_{uv}$ is integral for each $uv \in J$), for all $i \in Z$.*

**Lemma 4.2.** *Let $J \subseteq E$ be an overlapping clique. For every feasible solution $x$ of $(LP_0)$, we have $|ones(x) \cap J| \leq 1$. Furthermore, for every level $t \geq 2$ and every feasible solution $y$ of $\text{LAS}_t(LP_0)$, we have $y(J) \leq 1$.*

**Lemma 4.3.** *Let $w$ be a leaf of $T$, and let $u$ be an ancestor of $w$ such that every internal node (if any) of $P_{w,u}$ has exactly one child. Let $\hat{e} = vu$ be the tree-edge of $P_{w,u}$ that is incident to $u$ (possibly, $v = w$). Then, $\delta_E(\hat{e})$ is an overlapping clique. In particular, $\delta_E(w)$ is an overlapping clique. Moreover, we have $y(\delta_E(w)) = 1$ for any feasible solution $y \in \text{LAS}_t(LP_0)$ where $t \geq 2$.*

Recall that the matching polytope of the subgraph induced by the leaves, $G(L) = (L, E(L))$ is given by the following constraints:

$$x(\delta_{E(L)}(v)) \leq 1, \quad \forall v \in L$$
$$x(E(W)) \leq \frac{|W|-1}{2}, \quad \forall W \subseteq L, |W| \text{ odd}$$
$$x \geq 0$$

The next result is essentially the result on the matching polytope from the survey of Rothvoß, see [12, Lemma 13, Sec 3.3], translated to our setting.

**Lemma 4.4.** *Let $\epsilon > 0$, and let $y \in \text{LAS}_{\frac{1}{2\epsilon}+1}(LP_0)$. Then, $\frac{y|_{E(L)}}{1+\epsilon}$ is in the matching polytope of $G(L) = (L, E(L))$.*

**Lemma 4.5.** *Let $x$ be a feasible solution of $(LP_0)$ and let $s$ be a stem. Then $|ones(x) \cap \delta_E(s)| \leq 3$.*

*Proof.* Since $s$ is a stem, $s$ is incident to three tree-edges. Let $\hat{e}_1, \hat{e}_2, \hat{e}_3$ be the tree-edges incident with $s$. Clearly, $J_i = \delta_E(\hat{e}_i) \cap \delta_E(s)$ is an overlapping clique, for each $1 \leq i \leq 3$, and $\delta_E(s) = \cup_{1 \leq i \leq 3} J_i$. Then the result follows from Lemma 4.2. □

**Lemma 4.6.** *Let $t \geq 3$, and let $y$ be a feasible solution of $\text{LAS}_t(LP_0)$. Suppose that $s$ is a stem. Then:*

*(1) $y(twinlk(s)) \leq y(\delta_E^{\text{UP}}(s))$.*
*(2) For each bud $b$ in $T_s$ (if it exists), we have $y(buddylk(b)) \leq y(bs) + y(E(b, \mathcal{R}^{special}(b)))$.*

*Proof.* Let the leaves of $T_s$ be $b_0$ and $b_1$. Let $J = \delta_E(b_0) \cup \delta_E(b_1)$. Since $b_0, b_1$ are leaves, by Lemma 4.3, $\delta_E(b_0)$ and $\delta_E(b_1)$ are overlapping cliques. Thus, by Lemma 4.2, $|ones(x) \cap J| \leq 2$ for any feasible solution $x$ of $(LP_0)$. By Theorem 4.1, $y$ can be written as a convex combination $\sum_{i \in Z} \lambda_i x^i$ such that $x^i \in \text{LAS}_1(LP_0)$ and $x^i|_J$ is integral, $\forall i \in Z$.

---

[1]Formally speaking, $\text{LAS}_t(LP_0)$ is a set of vectors in $\mathbb{R}^{2^{[|E|]}}$. In what follows, we abuse the notation and take $\text{LAS}_t(LP_0)$ to be the projection on the subspace indexed by the singleton sets; thus, we take $\text{LAS}_t(LP_0)$ to be a set of vectors in $\mathbb{R}^{|E|}$.

**(1):** Consider statement (1) of the lemma. Observe that $twinlk(s) = b_0 b_1$. Since $b_0 b_1 \in J$, either $x^i(b_0 b_1) = 0$ or $x^i(b_0 b_1) = 1$, for all $i \in Z$. Let $Z_{(1)} = \{i \; : \; x^i(b_0 b_1) = 1\}$. Then $y(b_0 b_1) = \sum_{i \in Z_{(1)}} \lambda_i x^i(b_0 b_1)$. Consider $x^i$, where $i \in Z_{(1)}$. Let $\hat{e}_s$ be the tree-edge between $s$ and its parent. Then, $x^i(\delta_E(\hat{e}_s)) \geq 1$. Notice that every link in $\delta_E(\hat{e}_s)$ with positive $x^i$-value must have $s$ as its end in $T_s$; otherwise, if such a link has an end at some other node of $T_s$, then it will be overlapping with the link $b_0 b_1$, and we would have a violation of the overlapping constraints of $(LP_0)$. Thus, $x^i(\delta_E(\hat{e}_s)) = x^i(\delta_E(\hat{e}_s) \cap \delta_E(s))$. Hence, $x^i(\delta_E^{\text{UP}}(s)) = x^i(\delta_E(\hat{e}_s) \cap \delta_E(s)) = x^i(\delta_E(\hat{e}_s)) \geq 1 = x^i(b_0 b_1), \forall i \in Z_{(1)}$. Consequently, $y(b_0 b_1) = \sum_{i \in Z_{(1)}} \lambda_i x^i(b_0 b_1) \leq \sum_{i \in Z_{(1)}} \lambda_i x^i(\delta_E^{\text{UP}}(s)) \leq \sum_{i \in Z} \lambda_i x^i(\delta_E^{\text{UP}}(s)) = y(\delta_E^{\text{UP}}(s))$.

**(2):** Consider statement (2) of the lemma. W.l.o.g., let $b = b_0$ (the same argument applies for $b = b_1$). Let $buddylk(b_0) = b_1 b_2$. Since $b_1 b_2 \in J$, either $x^i(b_1 b_2) = 0$ or $x^i(b_1 b_2) = 1$, for all $i \in Z$. Let $Z_{(2)} = \{i \; : \; x^i(b_1 b_2) = 1\}$. Then $y(b_1 b_2) = \sum_{i \in Z_{(2)}} \lambda_i x^i(b_1 b_2)$.

Consider $x^i$, where $i \in Z_{(2)}$. Let $\ell = b_0 w$ be a link such that $x^i(\ell) = 1$; such a link must exist. If $w \neq s$ and $w$ is in $P_{b_1,b_2}$, then $\ell$ is overlapping with $b_1 b_2$, and we would have a violation of the overlapping constraints of $(LP_0)$. Hence, by Fact 2.3, $w \in \mathcal{R}^{\text{special}}(b_0) \cup \{s\}$. Then, $x^i(buddylk(b_0)) = 1 \leq x^i(\ell) \leq x^i(b_0 s) + x^i(E(b_0, \mathcal{R}^{\text{special}}(b_0)))$.

Consequently, $y(buddylk(b_0)) = \sum_{i \in Z_{(2)}} \lambda_i x^i(b_1 b_2) \leq \sum_{i \in Z_{(2)}} \lambda_i (x^i(b_0 s) + x^i(E(b_0, \mathcal{R}^{\text{special}}(b_0)))) \leq \sum_{i \in Z} \lambda_i (x^i(b_0 s) + x^i(E(b_0, \mathcal{R}^{\text{special}}(b_0)))) = y(b_0 s) + y(E(b_0, \mathcal{R}^{\text{special}}(b_0)))$. $\square$

For any stem $s$ and for any $x \in \mathbb{R}^E$, we define $slack_x(s)$ to be
$$\frac{1}{2}\Big(x(\delta_E^{\text{DN}}(s)) + \sum_{b \in L^{bud}(s)} (x(E(b, \mathcal{R}^{\text{special}}(b))) - x(buddylk(b)))\Big) + \frac{1}{2}\Big(x(\delta_E^{\text{UP}}(s)) - x(twinlk(s))\Big).$$

Lemma 4.6 implies the following fact.

**Fact 4.7.** *For any stem $s$ and for any feasible solution $y$ to $\text{LAS}_t(LP_0)$ where $t \geq 3$, we have $slack_y(s) \geq 0$.*

**Remark 4.8.** *Although the above fact assumes that the level $t$ is $\geq 3$, observe that $slack_x(s)$ is well defined for all levels $\geq 0$.*

## 5. Potential function

This section presents the potential function used by our algorithm. The potential function is based on a feasible solution $y$ to the Lasserre tightening of $(LP_0)$. It is essential to tighten $(LP_0)$ because the potential function is not valid for feasible solutions of $(LP_0)$. Possibly, our potential function may be obtained via weaker lift-and-project systems (e.g., Lovász-Schrijver, or Sherali-Adams). We prefer to apply the Lasserre tightening because our analysis in Sections 7–8 makes essential use of the decomposition theorem (Theorem 4.1), and this theorem is not known to hold for any weaker lift-and-project system; see Part I (Section 10).

Our potential function is defined via a subset of the leaves that is denoted by $\Lambda$. This subset is determined by the instance of TAP. Informally speaking, it consists of all the leaves of all occurrences of a particular type of subtree, called a bad 2-stem tree, see Section 6. Our potential function consists of two parts, a "preprocessing" part and a "normal" part. Our algorithm applies two preprocessing steps; the first preprocessing step contracts all occurrences of maximal bad 2-stem trees, and for this we have to "charge" the "preprocessing" part of our potential function.

We say that a set of leaves $\Lambda$ is *compatible* if the following holds:
- For every twin link and for every buddy link, either both ends of the link are in $\Lambda$ or no end of the link is in $\Lambda$; in other words, no twin link and no buddy link is present in $E(\Lambda, L - \Lambda)$.

In what follows, let $\Lambda \subseteq L$ denote a compatible set.

We denote the set of stems with both leaves in $\Lambda$ by $\mathcal{S}_\Lambda$. Similarly, let $L_\Lambda^{bud}$ denote the set of buds in $\Lambda$, and let $L_{(L-\Lambda)}^{bud}$ denote the set of buds in $L-\Lambda$. By the definition of a compatible set, the following fact holds.

**Fact 5.1.**

$$L_{(L-\Lambda)}^{bud} = \cup_{s \in \mathcal{S} - \mathcal{S}_\Lambda} L^{bud}(s),$$
$$E^{twinlk}(L-\Lambda) = \{twinlk(s) : s \in \mathcal{S} - \mathcal{S}_\Lambda\},$$
$$E^{buddylk}(L-\Lambda) = \cup_{s \in \mathcal{S} - \mathcal{S}_\Lambda} \{buddylk(b) : b \in L^{bud}(s)\}.$$

Let $\widehat{M}_{(L-\Lambda)}^{reg}$ denote a maximum matching of the subgraph $(L-\Lambda, E^{reg}(L-\Lambda))$, and let $\widehat{U}_{(L-\Lambda)}$ denote the set of nodes of this subgraph exposed by the matching $\widehat{M}_{(L-\Lambda)}^{reg}$; thus $\widehat{U}_{(L-\Lambda)} = (L-\Lambda) - \{v \in V : v \text{ is an end of some link } \ell \in \widehat{M}_{(L-\Lambda)}^{reg}\}$.

We mention that our potential function (the right-hand side of the inequality in Lemma 5.2 below) refers to the terms $\Lambda, \widehat{M}_{(L-\Lambda)}^{reg}, \widehat{U}_{(L-\Lambda)}$. Thus, when we use our potential function, we have to ensure that these terms have been defined already; we will "fix" our potential function by appropriately defining $\Lambda, \widehat{M}_{(L-\Lambda)}^{reg}, \widehat{U}_{(L-\Lambda)}$ in Section 6.

Given $\Lambda$ and $y \in \mathbb{R}^E$, we use $lbd_y(\Lambda)$ to denote the quantity

$$\frac{3}{2} y(E(\Lambda)) + \frac{1}{2} y(E(\Lambda, L-\Lambda)) + y(E(\Lambda, V-L)) + \frac{1}{2} \sum_{s \in \mathcal{S}_\Lambda} y(\delta_E(s));$$

this is one of the terms in our potential function; informally speaking, this is the main component of the "preprocessing" part of the potential function.

**Lemma 5.2.** *Let $\epsilon > 0$ be a constant, and let $t \geq \max\{\frac{1}{2\epsilon} + 1, 3\}$. Let $y \in \text{LAS}_t(LP_0)$. Then,*

$$\begin{aligned}(\frac{3}{2} + \epsilon) y(E) &\geq \frac{3}{2} |\widehat{M}_{(L-\Lambda)}^{reg}| + |\widehat{U}_{(L-\Lambda)}| + lbd_y(\Lambda) \\ &+ \frac{1}{2} y(E(V-L)) + \frac{1}{2} \sum_{v \in \mathcal{R}^{nonspcl}} y(\delta_E(v)) + \frac{1}{2} \sum_{v \in \mathcal{R}^{special}} y(E(v, V - L_{(L-\Lambda)}^{bud}(v))) + \sum_{s \in \mathcal{S} - \mathcal{S}_\Lambda} slack_y(s).\end{aligned}$$

*Moreover, every item on the right-hand side of the inequality is nonnegative.*

*Proof.* Observe that $y$ is feasible for $\text{LAS}_t(LP_0)$ for $t \geq 3$, hence, $y \geq 0$ and (by Fact 4.7) $slack_y(s) \geq 0, \forall s \in \mathcal{S}$. Hence, every item on the right-hand side of the inequality is nonnegative.

For each link $uw$, we distribute the value $\frac{3}{2} y_{uw}$ as follows:

- if both $u, w \in L$, then $uw$ keeps the value $\frac{3}{2} y_{uw}$;
- if both $u, w \in V - L$, then $uw$ keeps the value $\frac{1}{2} y_{uw}$ and each of the ends $u$ and $w$ gets value $\frac{1}{2} y_{uw}$;
- if only one of $u$ or $w$ is in $V - L$, then $uw$ keeps the value $y_{uw}$ and the end in $V - L$ gets value $\frac{1}{2} y_{uw}$.

(In other words, each node in $V - L$ borrows value $\frac{1}{2} y_{uw}$ from each link $uw$ incident to it, and the link keeps the remaining value; links $uw$ that are not incident to $V - L$ keep all of the value $\frac{3}{2} y_{uw}$.)

Thus, we have

$$\begin{aligned}\frac{3}{2}y(E) &= \frac{3}{2}y(E(\Lambda)) + \frac{3}{2}y(E(L-\Lambda)) + \frac{3}{2}y(E(\Lambda, L-\Lambda)) \\ &\quad + \frac{1}{2}y(E(V-L)) + y(E(V-L,\Lambda)) + y(E(V-L,L-\Lambda)) + \frac{1}{2}\sum_{v\in V-L}y(\delta_E(v)).\end{aligned}$$

Then we increase the coefficients of the twin links and buddy links in $E(L-\Lambda)$ from $\frac{3}{2}$ to 2 by borrowing the value $\frac{1}{2}y(E^{twin}(L-\Lambda)) + \frac{1}{2}y(E^{buddy}(L-\Lambda))$ from the last term above, and adding it to the term $\frac{3}{2}y(E(L-\Lambda))$. Thus, we replace the term $\frac{3}{2}y(E(L-\Lambda))$ by $2y(E(L-\Lambda)) - \frac{1}{2}y(E^{reg}(L-\Lambda))$, and we replace the last term by

$$\begin{aligned}&\frac{1}{2}\sum_{v\in V-L}y(\delta_E(v)) - \frac{1}{2}y(E^{twin}(L-\Lambda)) - \frac{1}{2}y(E^{buddy}(L-\Lambda)) \\ &= \frac{1}{2}\sum_{s\in\mathcal{S}_\Lambda}y(\delta_E(s)) + \frac{1}{2}\sum_{v\in\mathcal{R}^{nonspcl}}y(\delta_E(v)) + \frac{1}{2}\sum_{v\in\mathcal{R}^{special}}y(\delta_E(v)) + \frac{1}{2}\sum_{s\in\mathcal{S}-\mathcal{S}_\Lambda}(y(\delta_E^{DN}(s)) + y(\delta_E^{UP}(s))) \\ &\quad - \frac{1}{2}y(E^{twin}(L-\Lambda)) - \frac{1}{2}y(E^{buddy}(L-\Lambda)) \\ &= \frac{1}{2}\sum_{s\in\mathcal{S}_\Lambda}y(\delta_E(s)) + \frac{1}{2}\sum_{v\in\mathcal{R}^{nonspcl}}y(\delta_E(v)) + \frac{1}{2}\sum_{v\in\mathcal{R}^{special}}y(\delta_E(v)) + \frac{1}{2}\sum_{s\in\mathcal{S}-\mathcal{S}_\Lambda}y(\delta_E^{UP}(s)) - \frac{1}{2}y(E^{twin}(L-\Lambda)) \\ &\quad + \frac{1}{2}\sum_{s\in\mathcal{S}-\mathcal{S}_\Lambda}y(\delta_E^{DN}(s)) - \frac{1}{2}y(E^{buddy}(L-\Lambda)) \\ &\stackrel{(1)}{=} \frac{1}{2}\sum_{s\in\mathcal{S}_\Lambda}y(\delta_E(s)) + \frac{1}{2}\sum_{v\in\mathcal{R}^{nonspcl}}y(\delta_E(v)) + \Big(\frac{1}{2}\sum_{v\in\mathcal{R}^{special}}y(E(v,V-L^{bud}_{(L-\Lambda)}(v))) + \sum_{b\in L^{bud}_{(L-\Lambda)}}y(E(b,\mathcal{R}^{special}(b)))\Big) \\ &\quad + \frac{1}{2}\sum_{s\in\mathcal{S}-\mathcal{S}_\Lambda}y(\delta_E^{UP}(s)) - \frac{1}{2}y(E^{twin}(L-\Lambda)) + \frac{1}{2}\sum_{s\in\mathcal{S}-\mathcal{S}_\Lambda}y(\delta_E^{DN}(s)) - \frac{1}{2}y(E^{buddy}(L-\Lambda)) \\ &\stackrel{(2)}{=} \frac{1}{2}\sum_{s\in\mathcal{S}_\Lambda}y(\delta_E(s)) + \frac{1}{2}\sum_{v\in\mathcal{R}^{nonspcl}}y(\delta_E(v)) + \frac{1}{2}\sum_{v\in\mathcal{R}^{special}}y(E(v,V-L^{bud}_{(L-\Lambda)}(v))) \\ &\quad + \frac{1}{2}\sum_{s\in\mathcal{S}-\mathcal{S}_\Lambda}\Big(y(\delta_E^{DN}(s)) + \sum_{b\in L^{bud}(s)}(y(E(b,\mathcal{R}^{special}(b))) - y(\text{buddylk}(b)))\Big) \\ &\quad + \frac{1}{2}\sum_{s\in\mathcal{S}-\mathcal{S}_\Lambda}\Big(y(\delta_E^{UP}(s)) - y(\text{twinlk}(s))\Big) \\ &= \frac{1}{2}\sum_{s\in\mathcal{S}_\Lambda}y(\delta_E(s)) + \frac{1}{2}\sum_{v\in\mathcal{R}^{nonspcl}}y(\delta_E(v)) + \frac{1}{2}\sum_{v\in\mathcal{R}^{special}}y(E(v,V-L^{bud}_{(L-\Lambda)}(v))) + \sum_{s\in\mathcal{S}-\mathcal{S}_\Lambda}\text{slack}_y(s),\end{aligned}$$

where (2) follows from Fact 5.1 due to the assumption that $\Lambda$ is compatible, and (1) follows from the following equation:

$$\bigcup_{w\in\mathcal{R}^{special}} E(w, L^{bud}_{(L-\Lambda)}(w)) = \bigcup_{b\in L^{bud}_{(L-\Lambda)}} E(b, \mathcal{R}^{special}(b)).$$

This equation is based on some observations. First, the set on the right-hand side is clearly a subset of the set on the left-hand side. Conversely, consider a link $wa$ in the set on the left-hand side, where $w \in \mathcal{R}^{special}$ and $a \in L^{bud}_{(L-\Lambda)}(w)$. Since $w$ is an ancestor of $a$ and $w$ is not on the tree-path of the link $buddylk(a)$, by Fact 2.3, we have $w$ is in $\mathcal{R}^{special}(a)$, i.e., $wa \in E(a, \mathcal{R}^{special}(a))$. Hence, the link $wa$ belongs to the set on the right-hand side as well.

Thus, the expression for $\frac{3}{2}y(E)$ can be written as

$$= \frac{3}{2}y(E(\Lambda)) + \frac{3}{2}y(E(\Lambda, L-\Lambda)) + 2y(E(L-\Lambda)) - \frac{1}{2}y(E^{reg}(L-\Lambda))$$

$$+ \frac{1}{2}y(E(V-L)) + y(E(V-L, \Lambda)) + y(E(V-L, L-\Lambda))$$

$$+ \frac{1}{2}\sum_{s \in \mathcal{S}_\Lambda} y(\delta_E(s)) + \frac{1}{2}\sum_{v \in \mathcal{R}^{nonspcl}} y(\delta_E(v)) + \frac{1}{2}\sum_{v \in \mathcal{R}^{special}} y(E(v, V-L^{bud}_{(L-\Lambda)}(v))) + \sum_{s \in \mathcal{S}-\mathcal{S}_\Lambda} slack_y(s)$$

$$= \left(\frac{3}{2}y(E(\Lambda)) + \frac{1}{2}y(E(\Lambda, L-\Lambda)) + y(E(\Lambda, V-L)) + \frac{1}{2}\sum_{s \in \mathcal{S}_\Lambda} y(\delta_E(s))\right)$$

$$+ \left(y(E(L-\Lambda, \Lambda)) + 2y(E(L-\Lambda)) + y(E(L-\Lambda, V-L))\right)$$

$$+ \frac{1}{2}y(E(V-L)) + \frac{1}{2}\sum_{v \in \mathcal{R}^{nonspcl}} y(\delta_E(v)) + \frac{1}{2}\sum_{v \in \mathcal{R}^{special}} y(E(v, V-L^{bud}_{(L-\Lambda)}(v))) + \sum_{s \in \mathcal{S}-\mathcal{S}_\Lambda} slack_y(s)$$

$$- \frac{1}{2}y(E^{reg}(L-\Lambda))$$

$$= lbd_y(\Lambda) + \sum_{v \in L-\Lambda} y(\delta_E(v)) + \frac{1}{2}y(E(V-L))$$

$$+ \frac{1}{2}\sum_{v \in \mathcal{R}^{nonspcl}} y(\delta_E(v)) + \frac{1}{2}\sum_{v \in \mathcal{R}^{special}} y(E(v, V-L^{bud}_{(L-\Lambda)}(v))) + \sum_{s \in \mathcal{S}-\mathcal{S}_\Lambda} slack_y(s)$$

$$- \frac{1}{2}y(E^{reg}(L-\Lambda))$$

By Lemma 4.3, we have $y(\delta_E(v)) = 1$ for any $v \in L$. Hence, we replace the term $\sum_{v \in L-\Lambda} y(\delta_E(v))$ (the 2nd term in the displayed equation above) by $|L-\Lambda|$. Moreover, by Lemma 4.4, we have $\frac{y|_{E(L)}}{1+\epsilon}$ is in the matching polytope of $(L, E(L))$. Thus, $\frac{y|_{E^{reg}(L-\Lambda)}}{1+\epsilon}$ is in the matching polytope of $(L-\Lambda, E^{reg}(L-\Lambda))$, which implies $y(E^{reg}(L-\Lambda)) \leq (1+\epsilon)|\widehat{M}^{reg}_{(L-\Lambda)}|$. We derive the inequality (stated in the lemma) by replacing the term $-\frac{1}{2}y(E^{reg}(L-\Lambda))$ (the last term in the displayed equation above) by $-\frac{1}{2}(1+\epsilon)|\widehat{M}^{reg}_{(L-\Lambda)}|$. Now, observe that $|L-\Lambda| - \frac{1}{2}(1+\epsilon)|\widehat{M}^{reg}_{(L-\Lambda)}| = |\widehat{U}_{(L-\Lambda)}| + \frac{3}{2}|\widehat{M}^{reg}_{(L-\Lambda)}| - \frac{1}{2}(\epsilon)|\widehat{M}^{reg}_{(L-\Lambda)}|$, because $\widehat{U}_{(L-\Lambda)} = (L-\Lambda) - \{v \in V : v \text{ is an end of some link } \ell \in \widehat{M}^{reg}_{(L-\Lambda)}\}$. Note that $|\widehat{M}^{reg}_{(L-\Lambda)}| \leq \frac{1}{2}|L| = \frac{1}{2}\sum_{v \in L} y(\delta_E(v)) \leq y(E)$, hence, $-\frac{\epsilon}{2}|\widehat{M}^{reg}_{(L-\Lambda)}| \geq -\epsilon y(E)$. Thus, we get our potential function:

$$(\frac{3}{2}+\epsilon)y(E) \geq \frac{3}{2}|\widehat{M}^{reg}_{(L-\Lambda)}| + |\widehat{U}_{(L-\Lambda)}| + lbd_y(\Lambda) + \frac{1}{2}y(E(V-L))$$

$$+ \frac{1}{2}\sum_{v \in \mathcal{R}^{nonspcl}} y(\delta_E(v)) + \frac{1}{2}\sum_{v \in \mathcal{R}^{special}} y(E(v, V-L^{bud}_{(L-\Lambda)}(v))) + \sum_{s \in \mathcal{S}-\mathcal{S}_\Lambda} slack_y(s).$$

□

## 6. Algorithm and credits I: Preprocessing steps

We state our main result for (unweighted) TAP:

**Theorem 6.1.** *Let $\epsilon > 0$ be any (small) constant, and let $t \geq \max\{17, \frac{1}{2\epsilon} + 1\}$. Let $y$ denote an optimal solution of $\text{LAS}_t(LP_0)$. The integrality ratio of $\text{LAS}_t(LP_0)$ is $\leq \frac{3}{2} + \epsilon$. Moreover, there is a polynomial-time algorithm for TAP that finds a feasible solution (a set of links that covers the tree $T$) of size $\leq (\frac{3}{2} + \epsilon)y(E)$.*

For the rest of this paper, we let $y \in \mathbb{R}^E$ denote an optimal solution of $\text{LAS}_t(LP_0)$, where $t \geq \max\{17, 1 + \frac{1}{2\epsilon}\}$, where $\epsilon > 0$ is any constant. Our goal is to show that our algorithm finds a set of links that covers $T$ of size $\leq (\frac{3}{2} + \epsilon)y(E)$. We achieve this goal using our potential function (this is the right-hand side of the inequality in Lemma 5.2); we will "fix" the potential function below by defining the relevant terms (namely, $\Lambda, \widehat{M}^{reg}_{(L-\Lambda)}, \widehat{U}_{(L-\Lambda)}$). Recall from Part I (Section 6) that the potential function provides credit to the algorithm.

Also, recall from Part I (Section 6) that the combinatorial algorithm is a greedy-type iterative algorithm that makes a leaves-to-root scan over the tree $T$ and (incrementally) constructs a set of links $F$ that covers $T$. The algorithm starts with $F := \emptyset$, at each step it adds one or more links to $F$ (it never removes links from $F$), and at termination, it outputs a cover $F$ of $T$ whose size is $\leq$ the potential function.

The algorithm incurs a cost of one unit for every link that it picks, and it incurs a cost of one unit for each new compound node that it creates in the execution. The key to the analysis is to show that for each step, the cost incurred is compensated by a part of the credit.

We mention that (with one exception, see Lemma 8.3) the nodes or links that get contracted into a compound node are no longer relevant for the algorithm or the analysis. In particular, the credit (if any) of such nodes or links may be used at the step when they get contracted into a compound node, but after that step, any remaining credit of such nodes or links is not used at all.

For the current tree $T' = T/F$ and nodes $u, w$ of $T'$, let $P'_{u,w}$ denote the path of $T'$ between $u$ and $w$. By a *fitting cover* of a rooted tree $T'_v \subseteq T'$ we mean a set of links $J$ that covers all of the tree-edges of $T'_v$ but does not cover any other tree-edge; thus, we have $\cup_{uw \in J} P'_{u,w} = T'_v$.

**6.1. Semiclosed trees.** We recall the notion of a semiclosed tree w.r.t. an arbitrary matching. This notion is due to Even, et al., based on earlier work by Nagamochi [6]; also, see [5, Definition 2.3].

Let $T'_v$ be a rooted subtree of the current tree $T' = T/F$. Let $\bar{M}$ be an arbitrary matching of the leaf-to-leaf links. $T'_v$ is called *semiclosed* w.r.t. $\bar{M}$ if the following conditions hold:
  (i) Each link in $\bar{M}$ either has both ends in $T'_v$ or has no end in $T'_v$.
  (ii) Every link incident to an $\bar{M}$-exposed leaf of $T'_v$ has both ends in $T'_v$.

Let $\bar{M}(T'_v)$ denote the set of links in $\bar{M}$ that have both ends in $T'_v$.

We define
$$\Gamma(\bar{M}, T'_v) := \bar{M}(T'_v) \bigcup \{up(w)w \,:\, w \text{ is an } \bar{M}\text{-exposed leaf in } T'_v\};$$

thus, we associate a "basic link set" with the pair $\bar{M}, T'_v$. In general, the "basic link set" may not be a cover of $T'_v$.

By a *minimally semiclosed tree* $T'_v$ we mean that $T'_v$ is semiclosed but none of the proper rooted subtrees of $T'_v$ is semiclosed.

**Lemma 6.2** (Even, et al., [5]). *Let $T'_v$ be a minimally semiclosed tree w.r.t $\bar{M}$. Then $\Gamma(\bar{M}, T'_v)$ is a fitting cover of $T'_v$.*

**6.2. Maximum matching.** Our algorithm and analysis are based on a *maximum* matching of the leaf-to-leaf links that are neither twin links nor buddy links. Let $M$ denote one such matching; thus, $M$ is a maximum matching of the subgraph $(L, E^{reg}(L))$. By an *M-link* we mean a link that is in $M$. We denote the set of $M$-exposed leaf nodes by $U$.

The image of $M$ w.r.t. $T'$ is $\{u'w' : uw \in M, u' \neq w'\}$, where $u', w'$ denote the images (w.r.t. $T'$) of the original nodes $u, w$ of $T$ (i.e., we have $u' = u$ if $u$ is an original node of $T'$, otherwise, $u'$ denotes the compound node of $T'$ that contains $u$; $w'$ is defined similarly). We abuse the notation and use $M$ to denote both $M$ and its image w.r.t. $T'$, and by an $M$-link of $T'$ we mean the image (w.r.t. $T'$) of an original $M$-link.

For the rest of the paper, unless mentioned otherwise, a semiclosed tree means a subtree that is semiclosed w.r.t. the matching $M$.

6.3. **Bad 2-stem trees.** Let $T_v$ be a semiclosed tree rooted at $v$ that has exactly two stems $s_1, s_2$ and four leaves, where we denote the leaves of the tree $T_{s_i}$ by $u_i, w_i$ for $i = 1, 2$.

By a *leafy 3-cover* of $T_v$ we mean a set of three links $J$ such that $J$ covers $T_v$, one of the links in $J$ has one end in $T_v$ and one end in $L - L(T_v)$, and the other two links in $J$ have both ends in $T_v$.

We call $T_v$ a *bad 2-stem tree* if (i) one of the links in $E(\{u_1, w_1\}, \{u_2, w_2\})$ is in $M$, (ii) two of the leaves are $M$-exposed, (iii) one of the leaves is incident to all the links in $E(\{u_1, w_1\}, \{u_2, w_2\})$ (thus $E(\{u_1, w_1\}, \{u_2, w_2\})$ has one or two links), (iv) there exists a cover of $T_v$ of size three, and (v) there exists no leafy 3-cover of $T_v$. Let us fix the notation such that $w_1 w_2$ is the unique $M$-link in $E(L(T_v))$; thus, $u_1$ and $u_2$ are $M$-exposed (see Figure 4).

By a *maximal bad 2-stem tree* $T_v$ we mean a bad 2-stem tree that is not a proper subtree of another bad 2-stem tree. (Thus, any tree $T_q$ rooted at a proper ancestor $q$ of $v$ (if $q$ exists) must violate one of the conditions for being a bad 2-stem tree.)

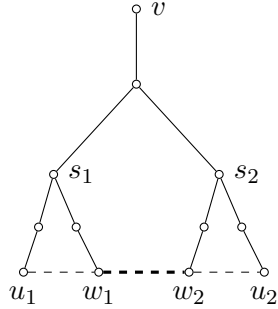

FIGURE 4. Illustration of a bad 2-stem tree. The dashed lines indicate links and the thick dashed line indicates an $M$-link.

Let $\mathcal{F}^{prep} = \{T_{v_1}, \ldots, T_{v_k}\}$ denote the set of maximal bad 2-stem trees of $T$; clearly, any two trees in $\mathcal{F}^{prep}$ are disjoint, by Property 2.1. We use $V(\mathcal{F}^{prep})$ to denote $\bigcup_{T_v \in \mathcal{F}^{prep}} V(T_v)$. We define $\Lambda = \bigcup_{T_v \in \mathcal{F}^{prep}} L(T_v)$, that is, $\Lambda$ consists of all the leaves of all the trees in $\mathcal{F}^{prep}$.

Since each bad 2-stem tree has a cover of size 3, the shadow-closed property implies that each bad 2-stem tree has a fitting cover of size $\leq 3$. Our algorithm applies a preprocessing step that contracts each tree $T_v \in \mathcal{F}^{prep}$ by a fitting cover of size $\leq 3$.

**Preprocessing step 1 ($\Lambda$-contraction):** For every tree $T_v \in \mathcal{F}^{prep}$, add a fitting cover of $T_v$ of size $\leq 3$ to $F$ and contract $T_v$ to a compound node.

The cost incurred for this step is $\leq 4|\mathcal{F}^{prep}|$, since each tree in $\mathcal{F}^{prep}$ incurs a cost of $\leq 3$ for its fitting cover and a cost of 1 for the resulting compound node.

This cost is charged to one part of our potential function, namely, it is charged to

$$lbd_y(\Lambda) + \frac{1}{2} \sum_{v \in \mathcal{R}^{nonspcl} \cap V(\mathcal{F}^{prep})} y(\delta_E(v)).$$

Lemma 6.6 below shows that this quantity is $\geq 4|\mathcal{F}^{prep}|$.

**Remark 6.3.** *The results in this section show that $\Lambda$-contraction is valid, in the sense that the algorithm has sufficient credits to pay for the cost of this step. Moreover, $\Lambda$-contraction is essential for the overall analysis (i.e., the algorithm cannot skip $\Lambda$-contraction), because the proof of Theorem 8.7 (Subcase 2.2) relies on $\Lambda$-contraction.*

**Lemma 6.4.** *Let $\mathcal{F}^{\text{prep}}$ and $\Lambda$ be as defined above. Then, we have*

$$lbd_y(\Lambda) + \frac{1}{2} \sum_{v \in \mathcal{R}^{\text{nonspcl}} \cap V(\mathcal{F}^{\text{prep}})} y(\delta_E(v))$$
$$\geq \sum_{T_v \in \mathcal{F}^{\text{prep}}} \left( \frac{3}{2} y(E(L(T_v))) + \frac{1}{2} y(E(L(T_v), L-L(T_v))) + y(E(L(T_v), V-L)) + \frac{1}{2} \sum_{u \in V(T_v)-L} y(\delta_E(u)) \right)$$

*Proof.* Recall that $lbd_y(\Lambda)$ denotes $\frac{3}{2} y(E(\Lambda)) + \frac{1}{2} y(E(\Lambda, L-\Lambda)) + y(E(\Lambda, V-L)) + \frac{1}{2} \sum_{s \in \mathcal{S}_\Lambda} y(\delta_E(s))$.

Observe that a bad 2-stem tree has no buds, so every node of such a tree is either a leaf, or a stem, or a node of $\mathcal{R}^{\text{nonspcl}}$. Hence,

$$(\mathcal{R}^{\text{nonspcl}} \cap V(\mathcal{F}^{\text{prep}})) \cup \mathcal{S}_\Lambda = \bigcup_{T_v \in \mathcal{F}^{\text{prep}}} (V(T_v)-L)$$

and so we have

$$\frac{1}{2} \sum_{v \in \mathcal{R}^{\text{nonspcl}} \cap V(\mathcal{F}^{\text{prep}})} y(\delta_E(v)) + \frac{1}{2} \sum_{s \in \mathcal{S}_\Lambda} y(\delta_E(s)) = \sum_{T_v \in \mathcal{F}^{\text{prep}}} \frac{1}{2} \sum_{u \in V(T_v)-L} y(\delta_E(u)).$$

We need the following claim to prove the lemma.

**Claim 6.5.** *Let $\Lambda_1, \Lambda_2, \ldots, \Lambda_k$ be a partition of $\Lambda$. Then, we have*

$$\frac{3}{2} y(E(\Lambda)) + \frac{1}{2} y(E(\Lambda, L-\Lambda)) + y(E(\Lambda, V-L))$$
$$\geq \frac{3}{2} \sum_{i=1}^{k} y(E(\Lambda_i)) + \frac{1}{2} \sum_{i=1}^{k} y(E(\Lambda_i, L-\Lambda_i)) + \sum_{i=1}^{k} y(E(\Lambda_i, V-L)).$$

We partition $\Lambda$ into the sets $\Lambda_i = L(T_{v_i})$, where $T_{v_1}, \ldots, T_{v_k}$ are the maximal bad 2-stem trees in $\mathcal{F}^{\text{prep}}$. Then, we have

$$lbd_y(\Lambda) + \frac{1}{2} \sum_{v \in \mathcal{R}^{\text{nonspcl}} \cap V(\mathcal{F}^{\text{prep}})} y(\delta_E(v))$$
$$= \frac{3}{2} y(E(\Lambda)) + \frac{1}{2} y(E(\Lambda, L-\Lambda)) + y(E(\Lambda, V-L)) + \sum_{T_v \in \mathcal{F}^{\text{prep}}} \frac{1}{2} \sum_{u \in V(T_v)-L} y(\delta_E(u))$$
$$\geq \sum_{T_v \in \mathcal{F}^{\text{prep}}} \left( \frac{3}{2} y(E(L(T_v))) + \frac{1}{2} y(E(L(T_v), L-L(T_v))) + y(E(L(T_v), V-L)) + \frac{1}{2} \sum_{u \in V(T_v)-L} y(\delta_E(u)) \right).$$

To prove the claim, observe that

$$\frac{3}{2}y(E(\Lambda)) + \frac{1}{2}y(E(\Lambda, L-\Lambda)) + y(E(\Lambda, V-L))$$

$$\geq \left(\frac{3}{2}\sum_{i=1}^{k} y(E(\Lambda_i, \Lambda_i)) + \frac{1}{2}\sum_{i=1}^{k}\sum_{j\neq i} y(E(\Lambda_i, \Lambda_j))\right) + \frac{1}{2}\sum_{i=1}^{k} y(E(\Lambda_i, L-\Lambda)) + \sum_{i=1}^{k} y(E(\Lambda_i, V-L))$$

$$= \frac{3}{2}\sum_{i=1}^{k} y(E(\Lambda_i)) + \frac{1}{2}\sum_{i=1}^{k} y(E(\Lambda_i, L-\Lambda_i)) + \sum_{i=1}^{k} y(E(\Lambda_i, V-L)).$$

□

**Lemma 6.6.** *Let $T_v$ be a bad 2-stem tree. Then we have*

$$\frac{3}{2}y\big(E(L(T_v))\big) + \frac{1}{2}y\big(E(L(T_v), L-L(T_v))\big) + y\big(E(L(T_v), V-L)\big) + \frac{1}{2}\sum_{u\in V(T_v)-L} y(\delta_E(u)) \geq 4.$$

*Hence, we have*

$$lbd_y(\Lambda) + \frac{1}{2}\sum_{v\in\mathcal{R}^{nonspcl}\cap V(\mathcal{F}^{prep})} y(\delta_E(v)) \geq 4|\mathcal{F}^{prep}|.$$

*Proof.* The second statement follows immediately from the first statement and Lemma 6.4. We focus on the first statement.

Let $s_1, s_2$ denote the two stems in $T_v$, and let $u_1, w_1$ (respectively, $u_2, w_2$) denote the two leaves in $T_{s_1}$ (respectively, $T_{s_2}$). Thus, $L(T_v) = \{u_1, w_1, u_2, w_2\}$. W.l.o.g. let $w_1w_2$ denote the unique $M$-link in $E(L(T_v))$; the two twin links $u_1w_1, u_2w_2$ are also in $E(L(T_v))$, and $E(L(T_v))$ may contain one other link incident to $w_1$ or $w_2$; there is no link between $u_1$ and $u_2$ (see Figure 4). Note that every link incident to $u_1$ or $u_2$ (the $M$-exposed leaves of $T_v$) must have both ends in $T_v$, since $T_v$ is semiclosed w.r.t. $M$.

Let $J$ denote the set of links incident to the leaves of $T_v$, thus, $J = \bigcup_{i=1,2}\big(\delta_E(u_i)\cup\delta_E(w_i)\big)$. For any feasible solution $x$ of $(LP_0)$, Lemmas 4.3 implies that $|ones(x)\cap J| \leq 4$. Hence, by Theorem 4.1, $y$ can be written as a convex combination $\sum_{i\in Z}\lambda_i x^i$ such that $x^i \in \text{Las}_3(LP_0)$ and $x^i|_J$ is integral, $\forall i \in Z$. (In what follows, we have to apply Lemma 4.6 to $x^i$, hence, $x^i$ must be feasible for level 3 (or higher) of the Lasserre tightening.)

Thus, it suffices to show that for any $i \in Z$, we have

$$\frac{3}{2}x^i\big(E(L(T_v))\big) + \frac{1}{2}x^i\big(E(L(T_v), L-L(T_v))\big) + x^i\big(E(L(T_v), V-L)\big) + \frac{1}{2}\sum_{u\in V(T_v)-L} x^i(\delta_E(u)) \geq 4;$$

let $\alpha$ denote the left-hand side of the above inequality.

Observe that every link in $J$ with positive $x^i$-value must have $x^i$-value one. Suppose that one of the links $\ell \in E(L(T_v))$ has positive $x^i$-value, thus $x^i(\ell) = 1$; then, we have $\frac{3}{2}x^i(E(L(T_v))) \geq \frac{3}{2}$, thus $\ell$ contributes value $\frac{3}{2}$ to $\alpha$. Also, if one of the links $\ell$ with one end in $L(T_v)$ and the other end at a non-leaf node of $T_v$ has positive $x^i$-value, then we have $x^i(E(L(T_v), V-L)) + \frac{1}{2}\sum_{u\in(V(T_v)-L)} x^i(\delta_E(u)) \geq 1 + \frac{1}{2} = \frac{3}{2}$, thus, $\ell$ contributes value $\frac{3}{2}$ to $\alpha$.

We complete the proof by case analysis, by considering the number of links with positive $x^i$-value such that one end is in $L(T_v)$ and the other end is not in $T_v$.

**Case 1.:** Suppose that there are no links with positive $x^i$-value such that one end is in $L(T_v)$ and the other end is not in $T_v$. Then, we focus on the number of links in $E(L(T_v))$ that have positive $x^i$-value; this number is zero, one, or two, because every link incident to a leaf has $x^i$-value zero or one, and moreover, $x^i(\delta_E(u)) = 1$ for each leaf $u$ of $T_v$, by $(LP_0)$ and Lemma 4.3. Thus, we have three subcases.

**Subcase 1.1.:** $x^i(E(L(T_v))) = 0$. Every link of $x^i$-value one incident to a leaf of $T_v$ has its other end at a non-leaf node of $T_v$, hence, each such link contributes $\frac{3}{2}$ to $\alpha$; we have four such links, so $\alpha \geq 6$.

**Subcase 1.2.:** $x^i(E(L(T_v))) = 1$. The link $\ell$ in $E(L(T_v))$ with $x^i$-value one contributes $\frac{3}{2}$ to $\alpha$. The two leaves in $L(T_v)$ that are not incident to $\ell$ each contribute $\frac{3}{2}$ to $\alpha$, because each is incident to a link of $x^i$-value one that has its other end at a non-leaf node of $T_v$, hence, $\alpha \geq \frac{9}{2}$.

**Subcase 1.3.:** $x^i(E(L(T_v))) = 2$. Since $x^i(\delta_E(u)) = 1$ for each leaf $u$ of $T_v$, the links in $E(L(T_v))$ with $x^i$-value one share no ends. By the definition of bad 2-stem trees, all the links in $E(\{u_1, w_1\}, \{u_2, w_2\})$ are incident to one of the four leaves. It follows that the twin links $u_1 w_1$ and $u_2 w_2$ both have $x^i$-value one, and each contributes $\frac{3}{2}$ to $\alpha$. Next, focus on the stems in $T_v$; by Lemma 4.6, $x^i(\delta_E^{\text{UP}}(s_j)) \geq x^i(u_j w_j) = 1$ for $j = 1, 2$. Thus, each stem contributes at least $\frac{1}{2}$ to $\alpha$ via the term $\frac{1}{2} \sum_{u \in (V(T_v) - L)} x^i(\delta_E(u))$. Hence, $\alpha \geq 2(\frac{3}{2}) + 2(\frac{1}{2}) = 4$.

**Case 2.:** Suppose that there is at least one link with positive $x^i$-value such that one end is in $L(T_v)$ and the other end is not in $T_v$. Again, we focus on the number of links in $E(L(T_v))$ that have positive $x^i$-value; it can be seen that this number is zero or one. (Note that every link incident to a leaf has $x^i$-value zero or one, and moreover, for every leaf $u$ of $T_v$, we have $x^i(\delta_E(u)) = 1$.) Thus, we have two subcases.

**Subcase 2.1.:** $x^i(E(L(T_v))) = 0$. Every link of $x^i$-value one incident to $u_1$ or $u_2$ (the $M$-exposed leaves of $T_v$) has its other end at a non-leaf node of $T_v$, hence, each such link contributes $\frac{3}{2}$ to $\alpha$; we have two such links. Every link of $x^i$-value one incident to $w_1$ or $w_2$ (the $M$-covered leaves of $T_v$) has its other end either in $V - L$, and so contributes 1 to $x^i(E(L(T_v), V - L))$, or else has its other end in $L - L(T_v)$, and so contributes $\frac{1}{2}$ to $\frac{1}{2} x^i(E(L(T_v), L - L(T_v)))$; again, we have two such links. Hence, $\alpha \geq 2(\frac{3}{2}) + 2(\frac{1}{2}) = 4$.

**Subcase 2.2.:** $x^i(E(L(T_v))) = 1$. Let $\ell_v$ denote a link of positive $x^i$-value such that one end is in $L(T_v)$ and the other end, denoted by $q$, is not in $T_v$. Note that $x^i(\ell_v) = 1$. The end of $\ell_v$ in $L(T_v)$ must be $M$-covered; w.l.o.g. assume that this node is $w_1$.

Note that $x^i(\delta_E(u)) = 1$ for each leaf $u$ of $T_v$. Since $T_v$ is a bad 2-stem tree, the links in $E(L(T_v))$ that are not incident to $w_1$ are the twin link $u_2 w_2$ and possibly the link $u_1 w_2$ (there is no link between $u_1, u_2$). Thus, either $x^i(u_2 w_2) = 1$ or $x^i(u_1 w_2) = 1$, and this contributes $\frac{3}{2}$ to $\alpha$ via the term $\frac{3}{2} x^i(E(L(T_v)))$. In either case, one of the $M$-exposed leaves $u_1$ or $u_2$ is incident to a link of $x^i$-value one that has its other end at a non-leaf node of $T_v$, and this contributes $\frac{3}{2}$ to $\alpha$ via the term $x^i(E(L(T_v), V - L)) + \frac{1}{2} \sum_{u \in (V(T_v) - L)} x^i(\delta_E(u))$.

We have two subcases, depending on whether $q$ (the end of $\ell_v$ in $V - V(T_v)$) is a leaf or not.

**Subcase 2.2.1.:** $q \notin L$. Then, the link $\ell_v = w_1 q$ contributes 1 to $\alpha$ via the term $x^i(E(L(T_v), V - L))$. Then, we have $\alpha \geq \frac{3}{2} + \frac{3}{2} + 1 = 4$.

**Subcase 2.2.2.:** $q \in L$. Then, the link $\ell_v = w_1 q$ contributes $\frac{1}{2}$ to $\alpha$ via the term $\frac{1}{2} x^i(E(L(T_v), L - L(T_v)))$.

Moreover, observe that we cannot have $x^i(u_1 w_2) = 1$, otherwise the three links $\ell_v = w_1 q, u_1 w_2, u_2 w_2$ form a leafy 3-cover of $T_v$ (see Figure 5(a)); recall that a bad 2-stem tree cannot have a leafy 3-cover.

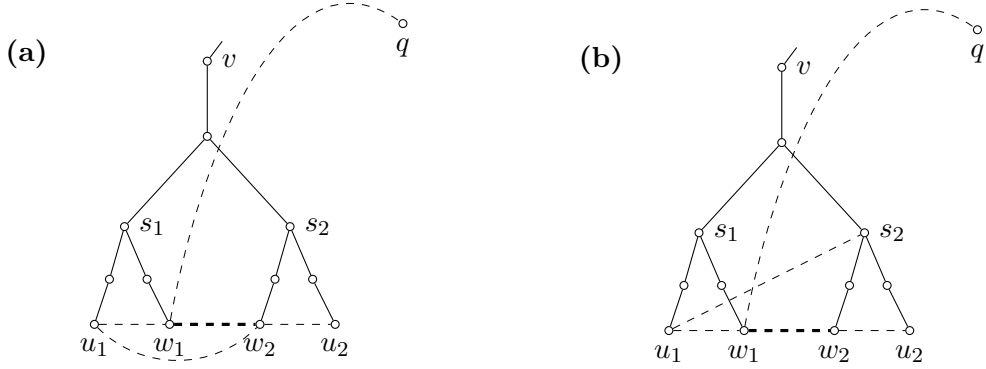

FIGURE 5. Illustration of leafy 3-covers of $T_v$ for Subcase 2.2.2 in the proof of Lemma 6.6. The dashed lines indicate links and the thick dashed lines indicate $M$-links.

Thus, $u_2w_2$ is the unique link in $E(L(T_v))$ with $x^i$-value one. Now, focus on the stem $s_2$; by Lemma 4.6, $x^i(\delta_E^{\text{UP}}(s_2)) \geq x^i(u_2w_2) = 1$. Thus, $\delta_E^{\text{UP}}(s_2)$ contributes at least $\frac{1}{2}$ to $\alpha$ via the term $\frac{1}{2}\sum_{u\in(V(T_v)-L)} x^i(\delta_E(u))$.

Finally, note that $u_1s_2$ is not present; otherwise, we would have a leafy 3-cover consisting of the (hypothetical) link $u_1s_2$ and the links $\ell_v = w_1q, u_2w_2$ (see Figure 5(b)). Hence, the link of $x^i$-value one incident with $u_1$ is not in $\delta_E^{\text{UP}}(s_2)$. Thus, we have $\alpha \geq \frac{3}{2} + \frac{3}{2} + \frac{1}{2} + \frac{1}{2} = 4$.

This completes the case analysis, and completes the proof. □

6.4. **Credit assignment for the algorithm and the preprocessing.** Recall that the algorithm starts with a number of credits equal to the potential function, namely, the right-hand side of the inequality in Lemma 5.2. In order to specify the potential function, we need to specify $\widehat{M}^{\text{reg}}_{(L-\Lambda)}, \widehat{U}_{(L-\Lambda)}$. We take $\widehat{M}^{\text{reg}}_{(L-\Lambda)}$ to be $M \cap E(L-\Lambda)$, i.e., the restriction of $M$ to the subgraph $(L-\Lambda, E^{\text{reg}}(L-\Lambda))$. It can be seen that $M \cap E(L-\Lambda) = M-E(\Lambda)$ is a maximum matching of the subgraph $(L-\Lambda, E^{\text{reg}}(L-\Lambda))$, as required by the definition of $\widehat{M}^{\text{reg}}_{(L-\Lambda)}$ in Section 5. We take $\widehat{U}_{(L-\Lambda)}$ to be $U \cap (L-\Lambda)$; again, this agrees with the definition of $\widehat{U}_{(L-\Lambda)}$ in Section 5.

The cost of $\Lambda$-contraction (Preprocessing step 1) is

$$\leq lbd_y(\Lambda) + \frac{1}{2} \sum_{u\in\mathcal{R}^{\text{nonspcl}}\cap V(\mathcal{F}^{\text{prep}})} y(\delta_E(u)),$$

where $\mathcal{F}^{\text{prep}}$ is defined above. We subtract this quantity from our potential function, and then add the credit of the compound nodes formed by contracting the trees $T_v \in \mathcal{F}^{\text{prep}}$, to get the remaining potential function (i.e., total credits available) for the rest of the execution. Clearly, the remaining potential function is

$$\frac{3}{2}|M \cap E(L-\Lambda)| + |U \cap (L-\Lambda)| + |\mathcal{F}^{\text{prep}}| +$$
$$\frac{1}{2}y(E(V-L)) + \frac{1}{2}\sum_{u\in(\mathcal{R}^{\text{nonspcl}}-V(\mathcal{F}^{\text{prep}}))} y(\delta_E(u)) + \frac{1}{2}\sum_{u\in\mathcal{R}^{\text{special}}} y(E(u, V-L^{\text{bud}}_{(L-\Lambda)}(u))) + \sum_{s\in\mathcal{S}-\mathcal{S}_\Lambda} \text{slack}_y(s)$$

By the *integral potential function* we mean the sum of the first three terms above (namely, $\frac{3}{2}|M \cap E(L-\Lambda)| + |U \cap (L-\Lambda)| + |\mathcal{F}^{\text{prep}}|$), and by the *fractional potential function* we mean the sum of the remaining terms (namely, the sum of the four terms that use $y$).

The following observation is useful for simplifying our notation.

**Fact 6.7.** *For the current tree $T'$ (after $\Lambda$-contraction), $U \cap (L-\Lambda)$ is the same as the set of $M$-exposed original leaves, and $M \cap E(L-\Lambda) = M(T')$.*

We start with the credit given by the integral potential function, and we maintain the following assignment of credits to the nodes of $T'$ and the links of $M(T') = M \cap E(L-\Lambda)$:

- every $M$-exposed original leaf has one credit,
- every compound node has one credit,
- every link of $M(T')$ has $\frac{3}{2}$ credit, and
- the root $r$ has one credit.

It can be seen that the integral potential function suffices for assigning credits to the tree $T'$ that results from $\Lambda$-contraction. (See Part I (Section 6) for a discussion on the unit credit for the root $r$.)

We define the *integral credit* of a set of links $J$ (w.r.t. $T'$) to be the sum of the credits of the $M$-links $pq$ such that $V(P'_{p,q}) \subseteq \bigcup_{uw \in J} V(P'_{u,w})$, plus the sum of the (integral) credits of the nodes in $\bigcup_{uw \in J} V(P'_{u,w})$, plus one if $r$ occurs as an original node in $\bigcup_{uw \in J} P'_{u,w}$.

In other words, the integral credit of $J$ is the sum of $\frac{3}{2}$ times the number of $M$-links $pq$ such that $V(P'_{p,q}) \subseteq \bigcup_{uw \in J} V(P'_{u,w})$, plus the number of compound nodes in $\bigcup_{uw \in J} V(P'_{u,w})$, plus the number of $M$-exposed original leaves in the same set, plus one if $r$ occurs as an original node in $\bigcup_{uw \in J} P'_{u,w}$. Informally speaking, when a step of the algorithm contracts all the links in $J$, then this amount of integral credit is available for this step (but not for subsequent steps).

Now, consider the fractional potential function. We use it to maintain an assignment of *fractional credits* to the (rooted) subtrees of $T'$. For any subtree $T'_v$ of $T'$, observe that $V(T'_v) \cap (\mathcal{R}^{\text{nonspcl}} - V(\mathcal{F}^{\text{prep}})) = V(T'_v) \cap \mathcal{R}^{\text{nonspcl}}$, because none of the original nodes in $V(\mathcal{F}^{\text{prep}})$ is present in $T'$ after the $\Lambda$-contraction; similarly, we have $V(T'_v) \cap (\mathcal{S} - \mathcal{S}_\Lambda) = V(T'_v) \cap \mathcal{S}$, because none of the stems in $\mathcal{S}_\Lambda$ is present in $T'$ after the $\Lambda$-contraction. For any subtree $T'_v$ of $T'$ and for any vector $x \in \mathbb{R}^E$, we use $\Phi(x, T'_v)$ to denote

$$\frac{1}{2} x(E((V-L) \cap V(T'_v), (V-L) \cap (V(T')-V(T'_v)))) +$$
$$\frac{1}{2} \sum_{u \in V(T'_v) \cap \mathcal{R}^{\text{nonspcl}}} x(\delta_E(u)) + \frac{1}{2} \sum_{u \in V(T'_v) \cap \mathcal{R}^{\text{special}}} x(E(u, V - L^{\text{bud}}_{(L-\Lambda)}(u))) + \sum_{s \in V(T'_v) \cap \mathcal{S}} \text{slack}_x(s).$$

The first term is defined on links that have both ends at original non-leaf nodes of $T'$, and moreover, have exactly one end in $T'_v$. Similarly, the last three terms denote credits that are assigned to original nodes in $T'_v$. We mention that the credits given by these four terms cannot be used two or more times by the algorithm. This holds because all credit (integral or fractional) associated with an original node of $T'$ can be used by a step that first contracts the original node to a new compound node; after that, none of the credit associated with this original node is available (any original node contained in a compound node is essentially "invisible" to the algorithm).

Informally speaking, $\Phi(y, T'_v)$ denotes the fractional credit of $T'_v$, that is, the part of the fractional potential function that is "owned" by $T'_v$. The fractional credit of $T'_v$ will be used together with its integral credit for contracting $T'_v$.

6.5. **Second preprocessing step.** We apply a second preprocessing step after the $\Lambda$-contraction and before the main loop of the algorithm. It turns out that the set of nodes contracted by this step is disjoint from the set of nodes contracted by the $\Lambda$-contraction.

Let $b_0$ be a bud, let $b_1 b_2 = buddylk(b_0)$, and let $s$ denote the stem such that $b_0$ is a leaf of $T_s$; w.l.o.g. let $b_1$ be the other leaf of $T_s$. Moreover, if $b_1$ is also a bud, then we assume w.l.o.g. that $up(b_0)$ is an ancestor of $up(b_1)$.

**Preprocessing step 2:** For each bud $b_0$ and the two ends $b_1, b_2$ of $buddylk(b_0)$ such that all three of $b_0, b_1, b_2$ are $M$-exposed, we contract the two links $up(b_0)b_0$ and $b_1 b_2$.

**Fact 6.8.** *Preprocessing step 2 has sufficient credits.*

*Proof.* Consider one "triple" consisting of a bud $b_0$ and the two ends $b_1, b_2$ of $buddylk(b_0)$ such that $b_0, b_1, b_2$ are $M$-exposed. Note that each of the $M$-exposed leaves $b_0, b_1, b_2$ has 1 unit of credit. Thus, we have sufficient credit for contracting two links and forming one compound node. □

**Remark 6.9.** *The results in this section show that Preprocessing step 2 is valid, in the sense that the algorithm has sufficient credits to pay for the cost of this step. Moreover, Preprocessing step 2 is essential for the overall analysis (i.e., the algorithm cannot skip this step), because the proof of Lemma 8.3 relies on this step.*

## 7. Algorithm and credits II: Overall algorithm

We present pseudo-code for the overall algorithm, after discussing some preliminaries. Also, we state and prove several assertions, i.e., we prove some basic properties maintained by the algorithm. These assertions are critical for the analysis in the next section.

7.1. **(Up-to-5) greedy contractions and assertions on $M$.** Recall that the integral credit of a set of links $J$ is given by the credits of the $M$-links $pq$ such that $V(P'_{p,q}) \subseteq \bigcup_{uw \in J} V(P'_{u,w})$, plus the sum of the (integral) credits of the nodes in $\bigcup_{uw \in J} V(P'_{u,w})$, plus one if $r$ occurs as an original node in $\bigcup_{uw \in J} P'_{u,w}$.

We define an *(up-to-5) greedy contraction* to be a contraction of a set of links $J$ such that

(i) $|J| \leq 5$;
(ii) the contraction of $J$ results in a single compound node, i.e., $\bigcup_{uw \in J} P'_{u,w}$ forms a connected graph;
(iii) the integral credit of $J$ is $\geq |J| + 1$.

Clearly, if an (up-to-5) greedy contraction is applicable, then, in polynomial time, the algorithm can find a set of links $J$ that is eligible for this step (by examining all link sets of size $\leq 5$).

**Remark 7.1.** *Clearly, the algorithm has sufficient credit for every application of (up-to-5) greedy contraction. The overall analysis relies on repeatedly applying (up-to-5) greedy contractions until no such contractions are applicable. In fact, every lemma/theorem in Section 8 assumes that no (up-to-5) greedy contractions are applicable.*

The following assertion is similar to Lemma 6.3 of Part I.

**Lemma 7.2** (Assertions on $M$). *Suppose that no (up-to-5) greedy contractions are applicable. Then:*

(1) *For every $M$-link $uw$, every node in $P'_{u,w}$ is an original node. In particular, w.r.t. $T'$, both ends of each $M$-link are original leaf nodes.*
(2) *There exist no links between $M$-exposed leaves of $T'$.*

7.2. **Good semiclosed trees.** Let $T'_v$ be a (rooted) subtree of $T'$. Let $U(T'_v)$ denote the set of $M$-exposed leaves of $T'_v$ (including both original nodes and compound nodes). Let $C(T'_v)$ denote the set of compound non-leaf nodes of $T'_v$. Recall that $M(T'_v)$ denotes the set of $M$-links that have both ends in $T'_v$. Note that every node in $(V(T'_v) \cap \mathcal{R}) \bigcup (V(T'_v) \cap \mathcal{S})$ is an original node.

Recall that a semiclosed tree means a tree that is semiclosed w.r.t. the matching $M$, unless mentioned otherwise. After the preprocessing steps, whenever we mention semiclosed trees, we assume that no (up-to-5) greedy contractions are applicable (see Section 7.1). Then, Lemma 7.2(1) implies that $M$ is a set of leaf-to-leaf links w.r.t. the current tree $T'$. Hence, semiclosed trees (of $T'$ w.r.t. $M(T')$) are well defined.

We define the credit of a (rooted) subtree $T'_v$ of the current tree $T'$ to be the sum of the fractional credit of $T'_v$, namely, $\Phi(y, T'_v)$, and the integral credit of $T'_v$. The latter is given by the sum of the following terms: $\frac{3}{2}|M(T'_v)|$, the number of compound nodes in $T'_v$, the number of $M$-exposed original leaves in $T'_v$, and an additional one if the root $r$ is in $V(T'_v) \cap \mathcal{R}$. We call a semiclosed tree $T'_v$ good if its credit is $\geq |\Gamma(M, T'_v)| + 1$.

The next result can be proved using arguments similar to the arguments used for proving Lemma 6.4 of Part I (although the potential function and credits in Parts I and II are different).

**Lemma 7.3.** *Let $T'_v$ be a semiclosed tree. If at least one of the following conditions is satisfied, then $T'_v$ is good.*

- $T'_v = T'$
- $C(T'_v) \neq \emptyset$
- $|M(T'_v)| \geq 2$
- $\Phi(y, T'_v) \geq 1$
- $|M(T'_v)| = 1$ and $\Phi(y, T'_v) \geq \frac{1}{2}$.

7.3. **Summary of the algorithm.** We start with $F := \emptyset$ ($F$ is the set of links picked by the algorithm) and $T' := T$ ($T'$ is the current tree $T/F$).

> apply Preprocessing step 1 ($\Lambda$-contraction);
> apply Preprocessing step 2;
> **while** $T'$ *is not a single node* **do**
>     repeatedly apply (up-to-5) greedy contractions until no such contractions are applicable;
>     find a good semiclosed tree $T'_v$ with a fitting cover $J$ of size $|\Gamma(M, T'_v)|$ (Algorithm 2 in Section 8 presents the details for finding such a semiclosed tree);
>     add $J$ to $F$, contract $T'_v$ to a new compound node, update $T'$;
> **end**
> 
> **Algorithm 1:** Find an approximately optimal solution for TAP.

7.4. **Stem assertion of the algorithm.** This section presents a basic assertion that is important for our analysis.

Recall that the algorithm iteratively contracts a set of links such that the tree-paths associated with these links form a connected graph; the set of chosen links and their associated tree-paths are contracted to form a new compound node. When we say that a contraction *hits* a node $v$ of $T'$, we mean that a step of the execution contracts a set of links $J$ such that $v$ is in $\bigcup_{uw \in J} V(P'_{u,w})$; thus, at least one of the tree-edges incident with $v$ is covered by one of the links in $J$.

**Stem assertion:** *Let $s$ be a stem. The first contraction that hits a node of $T_s$ (during the execution of the algorithm) must hit $s$.*

**Property 7.4.** *The algorithm maintains the stem assertion.*

*Proof.* Let $s$ be a stem. First, consider the two preprocessing steps. Either all of the tree $T_s$ is contracted or none of the tree-edges of $T_s$ is contracted by Preprocessing step 1 ($\Lambda$-contraction). The same statement holds for Preprocessing step 2. Hence, the stem assertion is maintained by the two preprocessing steps.

Next, suppose that the first contraction (in the execution) that hits a node in $T_s$ is an (up-to-5) greedy contraction that contracts a set of links $J$. If one of the links $uw \in J$ has one end in $T_s$ and the other end not in $T_s$, then the tree path $P'_{u,w}$ (in the current tree w.r.t. the first contraction that hits a node of $T_s$) must contain $s$, hence, the stem assertion is maintained. The remaining possibility is that all links of $J$ have both ends in $T_s$. Then observe that the number of integral credits available in $T_s$ is either zero, one, or two, and, in the last case, both leaves of $T_s$ are $M$-exposed. (By definition of stem, the root $r$ is not in $T_s$, since $r \neq s$.) The greedy contraction of $J$ requires $|J| + 1$ integral credits. Thus, $|J| = 1$ and $J$ contains the twin link of $s$. Contracting the twin link clearly maintains the stem assertion.

Finally, suppose that the first contraction (in the execution) that hits a node in $T_s$ is the contraction of a good semiclosed tree $T'_v$. By Property 2.1, one of $T_s$, $T'_v$ is contained in the other. It is easily seen that no proper subtree of $T_s$ is a semiclosed tree. (Note that there is a sublink between each leaf of $T_s$ and $s$ via the twin link of $s$, hence, any semiclosed tree containing a leaf of $T_s$ that is $M$-exposed will contain $s$ too. Also, any semiclosed tree containing a leaf of $T_s$ that is $M$-covered will contain $s$ too.) The only remaining possibility is that $T'_v$ contains $T_s$; then, the contraction of $T'_v$ maintains the stem assertion. □

## 8. Analysis of the algorithm and deficient trees

This section has our main results. Informally speaking, the key result (Theorem 8.7) asserts the following: if a semiclosed tree $T'_v$ is not good, then either $T'_v$ is a deficient tree (defined below) or $T'_v$ is a particular type of tree that is easily bypassed by our analysis.

The analysis consists of two parts. In Section 8.2, using local integrality of feasible solutions to the Lasserre system, we show that all semiclosed trees are good, except for a few cases. The nontrivial cases give deficient trees. Section 8.3 shows how to handle deficient trees. This leads to an efficient algorithm for finding a good semiclosed tree together with a fitting cover of appropriate size.

**Deficient 3-leaf tree**: Suppose that $T'_v$ is a semiclosed tree with exactly three leaves $a, b_1, b_2$. Clearly, among the nodes $w$ of $T'_v$ either there is exactly one node with $\deg_{T'}(w) = 4$ or there are two nodes with degree 3 in $T'$. In the latter case, we denote these two nodes by $u$ and $q$; moreover, we fix the notation such that $u$ is an ancestor of $q$, and the leaf $b_1$ is not a descendant of $q$; thus, $a, b_2$ (but not $b_1$) are descendants of $q$. In the former case, we denote by $u$ the unique node that is incident to four tree-edges. We call $T'_v$ a *deficient 3-leaf tree* (see Figure 6) if (i) the link $b_1 b_2$ is present and it is in $M(T')$, (ii) the link $ab_1$ is present, and (iii) there exists a link $b_2 w$ such that $w \in V(T') - V(T'_v)$.

Moreover, in the first case (with a unique node $u$ in $T'_v$ with $\deg_{T'}(u) = 4$), if conditions (i)–(iii) hold with both labelings $(b_1, b_2)$ and $(b_2, b_1)$ of the $M$-link, then we fix the notation such that $up(b_2)$ is an ancestor of $up(b_1)$. For both cases, we call $b_2$ the *ceiling leaf* of $T'_v$.

**Deficient 4-leaf tree:** Let $T'_v$ be a semi-closed tree with 4 leaves and exactly one stem node $s$ such that all nodes in $T'_v$ are original, $|M(T'_v)| = 1$, and the link in $M(T'_v)$ has exactly one end in $T'_s$. Let the four leaves of $T'_v$ be $a, b_1, b_2, c$, let the two leaves of $T'_s$ be $a, b_1$, and let the link in $M(T'_v)$ be $b_1 b_2$. Note that $c$ is the $M$-exposed leaf of $T'_v$ that is not in $T'_s$. Let $p$ be the least common ancestor of $s$ and $c$. Then, $T'_v$ is called a *deficient 4-leaf tree* (see Figure 7) if (i) $T'_p$ contains all leaves of $T'_v$, (ii) the link $cs$ is present, and (iii) there exists a link $b_2 w$ such that $w \in V(T') - V(T'_v)$. We call the link $cs$ the *latch* of $T'_v$.

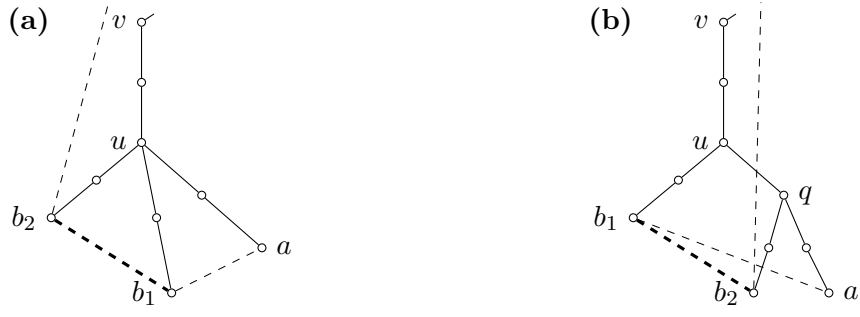

FIGURE 6. Illustration of deficient 3-leaf tree. The dashed lines indicate links and the thick dashed lines indicate $M$-links.

The contraction of the latch $cs$ in a deficient 4-leaf tree results in a deficient 3-leaf tree due to the presence of the links $ab_1, up(b_2)b_2$ (see Figure 6(a)). Let $b$ be the ceiling leaf of the resulting tree. Clearly, $up(b)$ is an ancestor of $up(b_2)$.

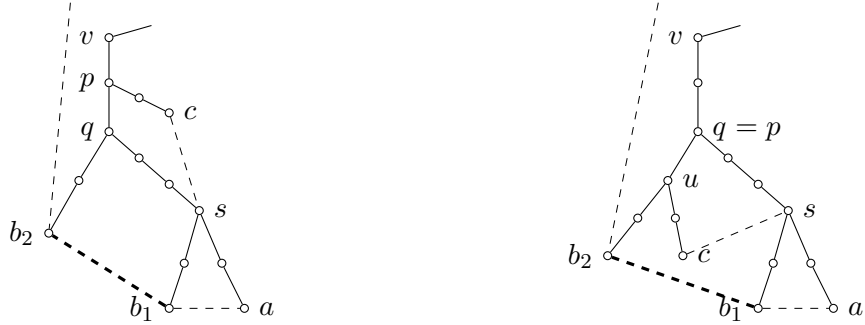

FIGURE 7. Illustration of deficient 4-leaf trees. The dashed lines indicate links and the thick dashed lines indicate $M$-links.

8.1. **Properties from assertions.** We start with an observation on compound leaf nodes. The next result (Lemma 8.2) states some properties pertaining to stem nodes and semiclosed trees; these properties are often used in the analysis of the algorithm. The proof relies on the stem assertion. After that we present a key lemma that pertains to some buds (Lemma 8.3); see the discussion preceding that lemma.

**Fact 8.1.** *If a compound leaf node contains a node $u \in V(T)$, then that compound node contains $T_u$.*

**Lemma 8.2.** *Suppose that no (up-to-5) greedy contractions are applicable. Let $T'_v$ be a semiclosed tree with $C(T'_v) = \emptyset$.*
  (1) *If $T'_v$ contains a stem node $s$, i.e., $s \in \mathcal{S} \cap V(T'_v)$, then every node in $T'_s$ is original.*
  (2) *In the original tree $T$, suppose that $s$ is a stem node, and $w$ is a leaf of $T_s$. If $w$ is contained in a compound node $\langle c \rangle$ that is an $M$-exposed leaf of $T'_v$, then all nodes of $T_s$ are contained in $\langle c \rangle$.*
  (3) *In the original tree $T$, suppose that $s$ is a stem node, and $w$ is a leaf of $T_s$. If $w$ is an original $M$-exposed leaf of $T'_v$, then $s$ is an original node of $T'_v$.*

*Proof.* The first part follows from the stem assertion.

Consider the second part. By the stem assertion, and the fact that $w$ is contained in $\langle c \rangle$, it can be seen that $s$ is contained in some compound node. If $s$ is contained in $\langle c \rangle$, then the proof is done by Fact 8.1. Now, suppose that $s$ is contained in a different compound node $\langle a \rangle$. Then, there exists a link between $\langle c \rangle$ and $\langle a \rangle$ (w.r.t. $T'$), because the link $ws$ is present in $E$ (the input) since $ws$ is a sublink of $twinlk(s)$. It can be seen that $T'_v$ contains $\langle a \rangle$ as a leaf, because $T'_v$ is semiclosed, $C(T'_v) = \emptyset$, $\langle c \rangle$ is an $M$-exposed leaf of $T'_v$, and the link between $\langle c \rangle$ and $\langle a \rangle$ is present. This gives a contradiction because the compound leaf node $\langle a \rangle$ contains $s$ so it contains $T_s$ (by Fact 8.1), hence, $\langle c \rangle$ cannot contain $w$.

The third part follows from arguments similar to that used for the second part; we give a sketch. Suppose that $s$ is contained in a compound node $\langle a \rangle$. If $\langle a \rangle$ is a leaf of $T'$, then $\langle a \rangle$ would contain the subtree $T_s$ (by Fact 8.1), and this would contradict the fact that the leaf $w$ is an original node. Since $C(T'_v) = \emptyset$, $\langle a \rangle$ cannot be a non-leaf node of $T'_v$. Thus, $T'_v$ contains $w$, but it does not contain $\langle a \rangle$. This contradicts the fact that $T'_v$ is semiclosed, because there is a link between the $M$-exposed leaf $w$ of $T'_v$ and $\langle a \rangle$ (due to the sublink $ws$ of $twinlk(s)$). □

The next lemma addresses a subtle point in our analysis. Consider a semiclosed tree $T'_v$ and its fractional credit $\Phi(y, T'_v)$, and observe that the links between a bud $b$ in $T'_v$ and a node in $V(T'_v) \cap \mathcal{R}^{special}(b)$ do not contribute to the fractional credit; see the third term in the definition of $\Phi(y, T'_v)$ (the last displayed equation in Section 6.4). Nevertheless, under appropriate conditions, we can exhibit fractional credits even when such links are present (see Lemma 8.5 and its proof). In order to do this, we have to focus on buds contained in $M$-exposed compound leaves such that there exists a link between the bud and a node "outside" the compound leaf (this is one part of the analysis that examines an original node even after it has been contracted into a compound node). The next lemma gives a tool for "counting" such buds, and the lemma is used twice in the subsequent analysis. The first use is in the proof of Lemma 8.4 to show that a semiclosed tree $T'_v$ with $|M(T'_v)| \leq 1$ has $O(1)$ such buds, and the second use is in the proof of Lemma 8.6 to show that a semiclosed tree $T'_v$ with $M(T'_v) = \emptyset$ has no such buds.

**Lemma 8.3.** *Suppose that no (up-to-5) greedy contractions are applicable. Let $T'_v$ be a semiclosed tree with $C(T'_v) = \emptyset$. Let $b_0$ be a bud that is contained in an $M$-exposed compound leaf $\langle c \rangle$ of $T'_v$ such that there exists a link $\ell$ incident to $b_0$ that has exactly one end (namely, $b_0$) in $\langle c \rangle$. Let $buddylk(b_0) = b_1 b_2$, and let $b_1 b_0$ be the twin link (w.r.t. $T$) incident to $b_0$. Then, $b_2$ is an original leaf node of $T'_v$ and $b_2$ is incident to a link of $M(T'_v)$.*

*Proof.* Let the original ends of $\ell$ be $b_0, w$; thus, $w$ is not contained in $\langle c \rangle$. Let $s$ be the stem associated with $b_0$; note that the leaves of $T_s$ are $b_0, b_1$. By Lemma 8.2(2), $T_s$ is contained in $\langle c \rangle$. Since $b_0$ is contained in an $M$-exposed leaf $\langle c \rangle$ of $T'_v$, $C(T'_v) = \emptyset$, and $T'_v$ is semiclosed, either $up(b_0)$ is contained in $\langle c \rangle$, or $up(b_0)$ is an original non-leaf node of $T'_v$. Clearly, $\langle c \rangle$ cannot contain $up(b_0)$, otherwise, $\langle c \rangle$ would contain $T_{up(b_0)}$ (by Fact 8.1), hence, the link $b_0 w$ would be contained in $\langle c \rangle$. Thus, $up(b_0)$ is an original non-leaf node of $T'_v$.

Consider the node $b_2$. Suppose that $b_2$ is an original leaf node of $T'$. Then, $b_2$ is in $T'_v$ (since $T'_v$ is semiclosed), and $b_2$ is $M$-covered (by Lemma 7.2(2) and the existence of the link $b_1 b_2$). Moreover, both ends of the $M$-link incident to $b_2$ are in $T'_v$ (since $T'_v$ is semiclosed). Thus, the lemma holds if $b_2$ is an original leaf node of $T'$.

Now, we may assume that $b_2$ is contained in a compound node $\langle a \rangle$. The hypotheses of the lemma imply that $\langle a \rangle$ is a leaf in $T'_v$. We must have $\langle a \rangle = \langle c \rangle$. Otherwise, either $\langle a \rangle$ is $M$-covered, contradicting Lemma 7.2(1), or $\langle a \rangle$ is $M$-exposed, contradicting Lemma 7.2(2) because the link $b_1 b_2$ implies a link between the $M$-exposed leaves $\langle a \rangle$, $\langle c \rangle$. We conclude that $\langle c \rangle$ contains all three of $b_0, b_1, b_2$. In what follows, we use this to derive a contradiction to the fact that $up(b_0)$ is an original non-leaf node in $T'_v$.

By Preprocessing step 2, one of $b_0, b_1, b_2$ is $M$-covered w.r.t. $T$, otherwise, the preprocessing step contracts two links to form a compound node that contains $up(b_0)$. We have two cases, depending on the $M$-links incident to $b_0, b_1, b_2$ w.r.t. $T$.

**Case 1:** There exists an $M$-link w.r.t. $T$ that has one end $b \in \{b_0, b_1, b_2\}$ and the other end $u \in L-\{b_0, b_1, b_2\}$. If $\langle c \rangle$ does not contain $u$, then we get a contradiction to Lemma 7.2(1) ($M$-link incident to compound node). Otherwise, $\langle c \rangle$ contains $u$, hence, $\langle c \rangle$ contains $P_{b,u}$; note that $up(b_0)$ is in $P_{b,u}$, so $\langle c \rangle$ contains $up(b_0)$. This is a contradiction.

**Case 2:** Each $M$-link w.r.t. $T$ incident to one of $b_0, b_1, b_2$ has both ends in $\{b_0, b_1, b_2\}$. There is only one such $M$-link, namely, $b_0 b_2$, since the twin link $b_0 b_1$ and the buddy link $b_1 b_2$ cannot be in $M$. Then $b_0$ is an $M$-covered bud w.r.t. $T$ (see Figure 8).

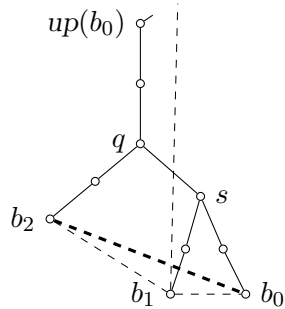

FIGURE 8. Illustration of an $M$-covered bud w.r.t. $T$.

We may assume that $b_0 b_2 \in M$ and $b_1$ is $M$-exposed. Suppose that $up(b_1)$ is a descendent of $up(b_0)$; then, $b_1$ is a bud; this follows from the definition of bud (see Section 2). Then, we get a contradiction since $b_0 b_2 \in M$ and $b_0 b_2 = buddylk(b_1)$ ($M$ has no buddy links). Hence, $up(b_1)$ is a proper ancestor of $up(b_0)$. Thus, $up(b_0) \neq r$, and no (rooted) subtree of $T_{up(b_0)}$ is semiclosed since $b_1$ is $M$-exposed.

The following claim gives a contradiction and completes the proof of this lemma.

> **Claim.** Suppose that $b_0 b_2$ is an $M$-link and $b_1$ is $M$-exposed. Then, the first contraction that hits a node of $T_{up(b_0)}$ (during the execution of the algorithm) must hit $up(b_0)$.

First, consider the two preprocessing steps. Preprocessing step 1 contracts all bad 2-stem trees. Note that no leaf of a bad 2-stem tree is a bud, hence, by Property 2.1, $T_{up(b_0)}$ is disjoint from every bad 2-stem tree. Preprocessing step 2 applies when a bud $b$ and the two ends of $buddylk(b)$ are all $M$-exposed. Since $b_0$ is $M$-covered, Preprocessing step 2 does not apply to $b_0, b_1, b_2$. Thus, the contractions in the preprocessing steps do not hit any node of $T_{up(b_0)}$.

Next, suppose that the first contraction (in the execution) that hits a node in $T_{up(b_0)}$ is an (up-to-5) greedy contraction that contracts a set of links $J$. If one of the links $uw \in J$ has one end in $T_{up(b_0)}$ and the other end not in $T_{up(b_0)}$, then the contraction must hit $up(b_0)$, because the tree path $P'_{u,w}$ (in the current tree at that contraction) contains $up(b_0)$. Thus, the claim holds. The remaining possibility is that all links of $J$ have both ends in $T_{up(b_0)}$. But observe that the number of integral credits available in $T_{up(b_0)}$ (in the current tree at that contraction) is exactly $\frac{5}{2}$ (due to one $M$-link and one $M$-exposed node), and this does not suffice for contracting two or more links; moreover, no single-link contraction applies because no singleton set $J$ has integral credit $> \frac{3}{2}$.

Finally, suppose that the first contraction (in the execution) that hits a node in $T_{up(b_0)}$ is the contraction of a good semiclosed tree $T'_w$. By Property 2.1, one of $T_{up(b_0)}$, $T'_w$ is contained in the other. Since no (rooted) subtree of $T_{up(b_0)}$ is semiclosed, it follows that $T'_w$ contains $T_{up(b_0)}$. Then, the contraction of $T'_w$ hits $up(b_0)$. This proves the claim, and completes the proof of the lemma.

□

8.2. **Most semiclosed trees are good.** Let $T'_v$ be a semiclosed tree. Let $L^{matched}(T'_v)$ denote the set of $M$-covered leaves of $T'_v$.

We call a compound node $\langle c \rangle$ of $T'_v$ *open* if it is an $M$-exposed leaf of $T'_v$, and moreover, $\langle c \rangle$ contains a bud $b_0$ such that there exists an original link between $b_0$ and an original node that is not contained in $\langle c \rangle$. We call such a bud an *open* bud (it must be contained in an open compound node and one of the original links incident to the bud is not contained in the compound node). Let $\mathcal{B}^{comp}(T'_v)$ denote the set of open buds of $T'_v$. Let $\mathcal{B}^{orig}(T'_v)$ denote the set of original nodes of $T'_v$ that are $M$-exposed buds. Let $\mathcal{B}(T'_v) = \mathcal{B}^{orig}(T'_v) \cup \mathcal{B}^{comp}(T'_v)$.

**Lemma 8.4.** *Suppose that no (up-to-5) greedy contractions are applicable. Let $T'_v$ be a semiclosed tree with $C(T'_v) = \emptyset$, $|M(T'_v)| \leq 1$ and $|\mathcal{S} \cap V(T'_v)| \leq 2$.*

*Then, $y$ can be written as a convex combination $\sum_{i \in Z} \lambda_i x^i$ such that $x^i \in \text{LAS}_3(LP_0)$ and $x^i|_J$ is integral, where $J = \{\ell \in \delta_E(u) : u \in (\mathcal{S} \cap V(T'_v)) \cup L^{matched}(T'_v) \cup \mathcal{B}(T'_v)\}$.*

*Proof.* We claim that $|ones(x) \cap J| \leq 14$ for any feasible solution $x$ of $(LP_0)$. The conclusion (of the lemma) follows easily from Theorem 4.1 and this claim.

To prove the claim, observe that $|\mathcal{S} \cap V(T'_v)| \leq 2$ and $|L^{matched}(T'_v)| = 2|M(T'_v)| \leq 2$, hence, by Lemmas 4.3, 4.5, we have

$$|ones(x) \cap \{\ell \in \delta_E(u) : u \in \mathcal{S} \cap V(T'_v)\}| \leq 6, \quad |ones(x) \cap \{\ell \in \delta_E(u) : u \in L^{matched}(T'_v)\}| \leq 2;$$

moreover, for each $u \in \mathcal{B}(T'_v)$, we have $|ones(x) \cap \delta_E(u)| \leq 1$. Thus, to complete the proof of the claim, we have to show that $|\mathcal{B}(T'_v)| \leq 6$.

First, consider any bud $b_0 \in \mathcal{B}^{orig}(T'_v)$ and its associated stem $s$. By Lemma 8.2(3), $s$ is an original node of $T'_v$, and so, $s \in \mathcal{S} \cap V(T'_v)$. By Lemma 8.2(1), each leaf of $T'_s$ is an original node in $T'$. Let $b_1$ be the leaf of $T'_s$ other than $b_0$. By Lemma 7.2(2), $b_1$ is $M$-covered (otherwise, the twin link $b_0 b_1$ connects two $M$-exposed leaves). Hence, $b_1 \notin \mathcal{B}^{orig}(T'_v)$. It follows that each stem in $\mathcal{S} \cap V(T'_v)$ has at most one leaf in $\mathcal{B}^{orig}(T'_v)$. Since $|\mathcal{S} \cap V(T'_v)| \leq 2$, we have $|\mathcal{B}^{orig}(T'_v)| \leq 2$.

Now, consider one of the buds $b_0 \in \mathcal{B}^{comp}(T'_v)$; $b_0$ is contained in some open compound node and one of the original links incident to $b_0$ has its other end "outside" this compound node. Then, by Lemma 8.3, the buddy link $buddylk(b_0)$ shares an end with an $M$-link in $M(T'_v)$. Since $|M(T'_v)| \leq 1$ and each end of an $M$-link can be incident to $\leq 2$ buddy links, we have $|\mathcal{B}^{comp}(T'_v)| \leq 4$. Therefore, $|\mathcal{B}(T'_v)| = |\mathcal{B}^{orig}(T'_v)| + |\mathcal{B}^{comp}(T'_v)| \leq 6$. This proves the claim. □

**Lemma 8.5.** *Suppose that no (up-to-5) greedy contractions are applicable. Let $T'_v$ be a semiclosed tree with $C(T'_v) = \emptyset$. Let $J$ be a set of links that each have at least one end in $T'_v$ and no end in $L^{matched}(T'_v) \cup (\mathcal{S} \cap V(T'_v))$. Let $x$ be a feasible solution of $\text{LAS}_3(LP_0)$ such that $x$ is integral on the links $\{\delta_E(u) : u \in \mathcal{B}(T'_v)\}$. Then,*

$$\Phi(x, T'_v) \geq \min\{\frac{1}{2}, \frac{1}{2} x(J)\}.$$

*Proof.* Let $g(x, T'_v)$ denote $\frac{1}{2} \sum_{u \in \mathcal{R}^{nonspcl} \cap V(T'_v)} x(\delta_E(u)) + \frac{1}{2} \sum_{u \in \mathcal{R}^{special} \cap V(T'_v)} x(E(u, V - L^{bud}(u)))$. By Fact 4.7, we have $\Phi(x, T'_v) \geq g(x, T'_v) + \sum_{s \in V(T'_v) \cap \mathcal{S}} slack_x(s) \geq g(x, T'_v)$. We will show that either each link $\ell \in J$ contributes $\frac{1}{2} x(\ell)$ to $g(x, T'_v)$, thereby ensuring $g(x, T'_v) \geq \frac{1}{2} x(J)$, or there exists a set of links that contribute $\frac{1}{2}$ to $g(x, T'_v)$.

Consider any link $\ell \in J$ with $x(\ell) > 0$. By Lemma 7.2(2), $\ell$ cannot have both ends at $M$-exposed leaves. Thus, $\ell$ has at least one end in $V(T'_v) \cap \mathcal{R}$, since it has no end in $L^{matched}(T'_v) \cup (\mathcal{S} \cap V(T'_v))$, $T'_v$ is semiclosed, and $C(T'_v) = \emptyset$. It is easily seen that, except for one case, $\ell$ contributes $\geq \frac{1}{2}x(\ell)$ to $g(x, T'_v)$.

The exceptional case occurs when $\ell$ has original ends $b_0$ and $w$, where $b_0 \in L^{bud}(w)$ and $w \in \mathcal{R}^{special}(b_0) \cap V(T'_v)$, see Fact 2.3. Since $J$ has no links incident to $L^{matched}(T'_v)$, it follows that $b_0 \in \mathcal{B}(T'_v)$ because either $b_0$ is an $M$-exposed original leaf of $T'_v$, or $b_0$ is contained in an $M$-exposed compound leaf of $T'_v$; in the latter case, $b_0$ is an open bud (w.r.t. $T'_v$). Then, we have $x(\ell) = 1$ because $x$ is integral on the links incident with nodes in $\mathcal{B}(T'_v)$. We claim that $g(x, T'_v) \geq \frac{1}{2}$. Let $\hat{e}$ denote the tree-edge between $w$ (the end of $\ell$ in $V(T'_v) \cap \mathcal{R}$) and its parent. Let $\delta^+(\hat{e})$ denote the set of links with positive $x$-value that cover $\hat{e}$. Clearly, $x(\delta^+(\hat{e})) \geq 1$. Consider any link $\ell_q = pq$ that is in $\delta^+(\hat{e})$, where $q$ is a descendent of $w$ (possibly, $q = w$) and $p$ is not in $T'_w$. If $q$ is a proper descendant of $w$, then (by the definition of $\mathcal{R}^{special}(b_0)$) $q$ is a descendant of the unique child of $w$ (see Figure 3), hence, $\ell$ and $\ell_q$ form an overlapping pair such that $x(\ell) + x(\ell_q) > 1$; this contradicts the overlapping constraints of $(LP_0)$. Hence, we have $q = w$. Clearly, $p$ (the other end of $\ell_q$) cannot belong to $L^{bud}(q) = L^{bud}(w)$ since $p$ is not a descendant of $q$. Hence, we have $g(x, T'_v) \geq \frac{1}{2}x(E(w, V - L^{bud}(w))) \geq \frac{1}{2}x(\delta^+(\hat{e})) \geq \frac{1}{2}$. This completes the proof. $\square$

Let $T'_v$ be a semiclosed tree. We construct an auxiliary graph in order to analyze the credits available in $T'_v$. We denote the auxiliary graph by $AG(T'_v)$. This is a bipartite multigraph, and the two sets in the node bipartition are denoted by $AS(T'_v)$ and $AU(T'_v)$. The first set consists of the $M$-covered leaves $L^{matched}(T'_v)$ and the stems $\mathcal{S} \cap V(T'_v)$. The second set contains an auxiliary node $\bar{v}$ (informally speaking, $\bar{v}$ represents the node set $V(T') - V(T'_v)$), as well as all the $M$-exposed leaves of $T'_v$, thus, $AU(T'_v) = \{\bar{v}\} \cup U(T'_v)$.

We define the edge set of $AG(T'_v)$ as follows: for every link $pq$ w.r.t. $T'$ with $p \in (\mathcal{S} \cap V(T'_v)) \cup L^{matched}(T'_v), q \in U(T'_v)$, the edge $pq$ is in $AG(T'_v)$, and for every link $pq$ w.r.t. $T'$ such that $p \in (\mathcal{S} \cap V(T'_v)) \cup L^{matched}(T'_v), q \in V(T') - V(T'_v)$, the edge $p\bar{v}$ is in $AG(T'_v)$. Thus, $AG(T'_v)$ is a multigraph (multiple copies of an edge may be present), and every edge in $AG(T'_v)$ corresponds to a link w.r.t. $T'$ (see Figure 9).

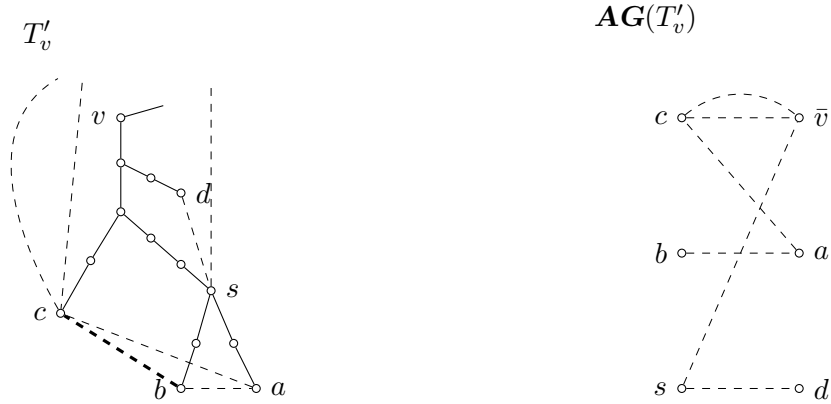

FIGURE 9. The left figure shows a semiclosed tree $T'_v$ where dashed lines indicate links and the thick dashed line indicates an $M$-link. The right figure shows the auxiliary graph $AG(T'_v)$ and its node bipartition $\{s, b, c\}, \{\bar{v}, a, d\}$.

In what follows, we may abuse the notation by not distinguishing between edges (sets of edges) of $AG(T'_v)$ and the corresponding links (sets of links) w.r.t. $T'$.

**Lemma 8.6.** *Suppose that no (up-to-5) greedy contractions are applicable. Let $T'_v$ be a semiclosed tree such that $T'_v \neq T'$, $C(T'_v) = \emptyset$, $|M(T'_v)| \leq 1$ and $|\mathcal{S} \cap V(T'_v)| \leq 2$.*

(1) *Suppose that $M(T'_v) = \emptyset$. Then for any feasible solution $x$ for $\text{LAS}_3(LP_0)$, we have $\Phi(x, T'_v) \geq 1$. Furthermore, $T'_v$ is good.*

(2) *Suppose that $|M(T'_v)| = 1$, and $|U(T'_v)| \geq |\mathcal{S} \cap V(T'_v)| + 1$. Moreover, suppose that $x$ is a feasible solution for $\text{LAS}_3(LP_0)$ such that $x|_J$ is integral and $\Phi(x, T'_v) < \frac{1}{2}$, where $J = \{\ell \in \delta_E(u) : u \in (\mathcal{S} \cap V(T'_v)) \cup L^{matched}(T'_v) \cup \mathcal{B}(T'_v)\}$.*

*Then, $|U(T'_v)| = |\mathcal{S} \cap V(T'_v)| + 1$. Moreover, the auxiliary graph has a perfect matching $AM(T'_v)$ such that the following conditions hold:*

(i) *$x(\ell) = 1$ for each link $\ell \in AM(T'_v)$,*
(ii) *the links of $AM(T'_v)$ cover $T'_v$,*
(iii) *for each stem node $s \in \mathcal{S} \cap V(T'_v)$, $twinlk(s)$ is in $AM(T'_v)$,*
(iv) *$AM(T'_v)$ has no links of the form $\bar{v}s$, where $s \in \mathcal{S} \cap V(T'_v)$.*

*Proof.* Let $\hat{e}_v$ denote the tree-edge between $v$ and its parent; $\hat{e}_v$ is well defined since $T'_v \neq T'$. Let $\bar{J} = \delta_E(\hat{e}_v) \cup (\cup_{u \in U(T'_v)} \delta_E(u))$. Then, $x(\bar{J}) = x(\delta_E(\hat{e}_v)) + \sum_{u \in U(T'_v)} x(\delta_E(u)) \geq 1 + |U(T'_v)|$; the equation holds because (i) $T'_v$ is semiclosed so none of the links in $\delta_E(\hat{e}_v)$ is incident to an $M$-exposed leaf of $T'_v$, and (ii) by Lemma 7.2(2), no link has both ends at $M$-exposed leaves; the inequality holds because $x(\delta_E(\hat{e})) \geq 1$ for every tree-edge $\hat{e}$.

Let $g(x, T'_v)$ denote $\frac{1}{2} \sum_{u \in \mathcal{R}^{nonspcl} \cap V(T'_v)} x(\delta_E(u)) + \frac{1}{2} \sum_{u \in \mathcal{R}^{special} \cap V(T'_v)} x(E(u, V - L^{bud}(u)))$.

By Fact 4.7, we have $\Phi(x, T'_v) \geq g(x, T'_v) + \sum_{s \in V(T'_v) \cap \mathcal{S}} slack_x(s) \geq g(x, T'_v)$.

**Case (1):** $M(T'_v) = \emptyset$. Clearly, $|U(T'_v)| \geq 1$, since $T'_v$ has at least one leaf.

Observe that $\frac{1}{2} x(\bar{J}) \geq \frac{1}{2}(1 + |U(T'_v)|) \geq 1$. We will show that every link $\ell \in \bar{J}$ contributes $\geq \frac{1}{2} x(\ell)$ to $g(x, T'_v)$.

First, we show that $\mathcal{S} \cap V(T'_v) = \emptyset$. Otherwise, consider any $s \in \mathcal{S} \cap V(T'_v)$; by Lemma 8.2(1), every node in $T'_s$ is original, and so the twin link of $s$ has both ends at $M$-exposed nodes (since $M(T'_v) = \emptyset$); this contradicts Lemma 7.2(2).

Next, we show that $\mathcal{B}(T'_v) = \emptyset$. Since $\mathcal{S} \cap V(T'_v) = \emptyset$, Lemma 8.2(3) implies that $\mathcal{B}^{orig}(T'_v) = \emptyset$ (if a bud $b_0 \in \mathcal{B}^{orig}(T'_v)$ is present as an original $M$-exposed leaf of $T'_v$, then its stem is present as an original node of $T'_v$). Also, we have $\mathcal{B}^{comp}(T'_v) = \emptyset$, otherwise, by Lemma 8.3, there exists a link in $M(T'_v)$ (this is impossible since $M(T'_v) = \emptyset$). Thus, $\mathcal{B}(T'_v) = \emptyset$.

Now, observe that every link $\ell \in \bar{J}$ has an end in $V(T'_v) \cap \mathcal{R}$, and hence, it contributes $\frac{1}{2} x(\ell)$ to $g(x, T'_v)$. Therefore, $\Phi(x, T'_v) \geq 1$. Since this inequality holds for every feasible solution $x$ of $\text{LAS}_3(LP_0)$, we have $\Phi(y, T'_v) \geq 1$. Hence, by Lemma 7.3, $T'_v$ is good.

**Case (2):** In this case, $|M(T'_v)| = 1$ and $|U(T'_v)| \geq |\mathcal{S} \cap V(T'_v)| + 1$. Moreover, $x|_J$ is integral, where $J = \{\ell \in \delta_E(u) : u \in (\mathcal{S} \cap V(T'_v)) \cup L^{matched}(T'_v) \cup \mathcal{B}(T'_v)\}$.

First, we show that $x(\delta_E(s)) \leq 1$ for each stem $s \in \mathcal{S} \cap V(T'_v)$. We use a contradiction argument. Suppose that $x(\delta_E(s)) > 1$. Let $\delta^+(s)$ denote the set of links of positive $x$-value incident to $s$. Note that every link $\ell$ in $\delta^+(s)$ has $x(\ell) = 1$ because $x|_J$ is integral. Since $x(\delta_E(s)) > 1$, we have $|\delta^+(s)| \geq 2$. This implies that each of the buddy links associated with the stem $s$ has $x$-value zero. To see this, note that the set of links in $\delta^+(s)$ covering each of the three tree-edges incident to $s$ is an overlapping clique, hence, at most one link of $\delta^+(s)$ belongs to one of these overlapping cliques (by the overlapping constraints of $(LP_0)$); moreover, each buddy link associated with $s$ is overlapping with the links that belongs to two of these overlapping cliques (see Figure 10); hence, if a buddy link $\ell$ associated with $s$ has $x(\ell) > 0$, then we get a violation (of an overlapping constraint of $(LP_0)$) for one of the three overlapping cliques. Since the buddy links associated with $s$ (if any) have $x$-value zero, we have $\Phi(x, T'_v) \geq \text{slack}_x(s) \geq \frac{1}{2}\Big(x(\delta_E(s)) - x(\textit{twinlk}(s))\Big) \geq \frac{1}{2}$, and this gives the desired contradiction.

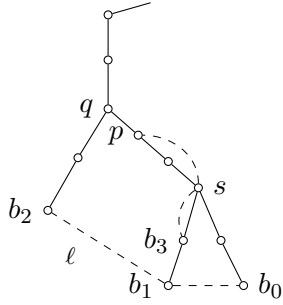

FIGURE 10. Illustration of a buddy link $\ell$ in the proof of Lemma 8.6. Note that $\ell$ is overlapping with links in both $\delta_E^{\text{UP}}(s)$ and $\delta_E(\hat{e}) \cap \delta_E(s)$, where $\hat{e}$ is the tree-edge between $b_3$ and $s$. For example, $\ell$ is overlapping with the links $sp$ and $sb_3$.

For any $M$-covered leaf $w$ in $T'_v$, by Lemma 7.2(1), $w$ is an original node, and moreover, we have $x(\delta_E(w)) \leq 1$, by Lemma 4.3. Thus, we have $x(\delta_E(w)) \leq 1$ for each node $w \in L^{\text{matched}}(T'_v) \cup (\mathcal{S} \cap V(T'_v)) = AS(T'_v)$.

Let $\widetilde{J} = \bar{J} - \bigcup_{w \in AS(T'_v)} \delta_E(w)$; this is the set of links in $\bar{J}$ but not in $AG(T'_v)$. We have $x(\widetilde{J}) < 1$; otherwise, by Lemma 8.5, we would have $\Phi(x, T'_v) \geq \min\{\frac{1}{2}, \frac{1}{2}x(\widetilde{J})\} \geq \frac{1}{2}$, which is a contradiction. Then, by the conditions in the hypothesis ($|M(T'_v)| = 1$ and $|U(T'_v)| \geq |\mathcal{S} \cap V(T'_v)| + 1$) and the fact that $|AU(T'_v)| = |U(T'_v)| + 1$, we have
$$1 > x(\widetilde{J}) = x(\bar{J}) - \sum_{w \in AS(T'_v)} x(\bar{J} \cap \delta_E(w)) \geq |AU(T'_v)| - |AS(T'_v)|$$
$$= |U(T'_v)| + 1 - 2|M(T'_v)| - |\mathcal{S} \cap V(T'_v)|$$
$$= |U(T'_v)| - (1 + |\mathcal{S} \cap V(T'_v)|) \geq 0.$$

Hence, we have $|U(T'_v)| = |\mathcal{S} \cap V(T'_v)| + 1$, and $|AS(T'_v)| = |AU(T'_v)|$. For each $w \in AS(T'_v)$, we have $x(\bar{J} \cap \delta_E(w)) = 1$ because $x(\ell)$ is integral for each link $\ell \in \bar{J} \cap \delta_E(w)$ (otherwise, $\sum_{w \in AS(T'_v)} x(\bar{J} \cap \delta_E(w)) \leq |AS(T'_v)| - 1$ implies that $x(\widetilde{J}) \geq |AU(T'_v)| - |AS(T'_v)| + 1 \geq 1$).

For each $w \in AS(T'_v)$, define $\ell_w$ to be the link in $\delta_E(w) \cap \bar{J}$ with $x(\ell_w) = 1$. We claim that these links form a perfect matching of the auxiliary graph, thus, $AM(T'_v) = \{l_w : w \in AS(T'_v)\}$. Otherwise, there exist two links $\ell_{w_1}, \ell_{w_2}$ (where $w_1, w_2 \in AS(T'_v)$) that are incident to the same node $u \in AU(T'_v)$. Then, $x(\bar{J}) \geq |AU(T'_v)| + 1$, which implies that $x(\widetilde{J}) = x(\bar{J}) - \sum_{w \in AS(T'_v)} x(\bar{J} \cap \delta_E(w)) \geq |AU(T'_v)| + 1 - |AS(T'_v)| \geq 1$. This contradicts the fact that $x(\widetilde{J}) < 1$. Our claim follows.

We claim that $AM(T'_v)$ covers $T'_v$. Otherwise, there exists a tree-edge $\hat{e}_0$ in $T'_v$ that is not covered by $AM(T'_v)$. Let $\delta^+(\hat{e}_0)$ denote the set of links of positive $x$-value in $\delta_E(\hat{e}_0)$. Then, we have $x(\delta^+(\hat{e}_0)) \geq 1$, and none of the links in $\delta^+(\hat{e}_0)$ is incident to $(\mathcal{S} \cap V(T'_v)) \cup L^{\text{matched}}(T'_v)$; the latter assertion holds because $x(\ell_w) = 1$ and $x(\delta_E(w)) \leq 1$ for every $w \in AS(T'_v)$, i.e., the nodes in $AS(T'_v)$ are already "saturated" by $AM(T'_v)$. Thus, Lemma 8.5 applies to $\delta^+(\hat{e}_0)$, and we have $\Phi(x, T'_v) \geq \min\{\frac{1}{2}, \frac{1}{2}\delta^+(\hat{e}_0)\} \geq \frac{1}{2}$, which is a contradiction. Our claim follows: $AM(T'_v)$ covers $T'_v$.

Additionally, we claim that $AM(T'_v)$ has no links between $\bar{v}$ and $\mathcal{S} \cap V(T'_v)$. By way of contradiction, assume that there exists a link $\ell = sq \in AM(T'_v)$ such that $s \in \mathcal{S} \cap V(T'_v)$ and $q \notin V(T'_v)$. Suppose that $q$ is an original non-leaf node. Then $sq$ is a link between two original non-leaf nodes with $x(sq) = 1$, hence, we have $\Phi(x, T'_v) \geq \frac{1}{2}$ due to Fact 4.7 and the following term in $\Phi(x, T'_v)$:

$$\frac{1}{2}x(E((V-L) \cap V(T'_v), (V-L) \cap (V(T') - V(T'_v)))).$$

This is a contradiction. Otherwise, if $q$ is a compound node or an original leaf, then we claim that an (up-to-5) greedy contraction applies, and this too gives a contradiction. Note that $|\mathcal{S} \cap V(T'_v)| \leq 2$ and $|M(T'_v)| = 1$, hence, $|AM(T'_v)| \leq 4$. The credit assigned to the leaves and matching links in $T'_v$ is $\frac{3}{2} + |U(T'_v)| = \frac{3}{2} + |\mathcal{S} \cap V(T'_v)| + 1 = |AM(T'_v)| + \frac{1}{2}$. If $q$ is a compound node or an original $M$-exposed leaf, then $q$ provides one additional credit, and this suffices for an (up-to-5) greedy contraction of the links of $AM(T'_v)$; otherwise, $q$ is an $M$-covered leaf, and the $M$-link $\ell_q$ incident to $q$ provides additional $\frac{3}{2}$ credit, and this suffices for an (up-to-5) greedy contraction of $\ell_q$ together with the links of $AM(T'_v)$. Our claim follows: $AM(T'_v)$ has no link of the form $\bar{v}s$, $s \in \mathcal{S} \cap V(T'_v)$.

Finally, we prove that $twinlk(s)$ is in $AM(T'_v)$ for each stem $s \in \mathcal{S} \cap V(T'_v)$. Consider any stem $s \in \mathcal{S} \cap V(T'_v)$. By Lemma 8.2(1), $T'_s$ is contained in $T'_v$ and every node in $T'_s$ is original. Let the leaves of $T'_s$ be $a, b_0$; thus, $twinlk(s) = ab_0$.

Lemma 7.2(2) implies that one end of the twin link $ab_0$ is $M$-covered. Since $x|_J$ is integral, and $J$ contains $ab_0$ (each link incident to $L^{matched}(T'_v)$ is in $J$), either $ab_0 \in AM(T'_v)$ or $x(ab_0) = 0$. We may assume that $x(ab_0) = 0$. Suppose that $\sum_{b \in L^{bud}(s)} x(buddylk(b)) = 0$, i.e., there does not exist a buddy link incident to $T'_s$ that has positive $x$-value. Consider the link $\ell_s$ in $AM(T'_v)$ incident with $s$; clearly, $x(\ell_s) = 1$. Thus, we have $\Phi(x, T'_v) \geq slack_x(s) \geq \frac{1}{2}x(\ell_s) = \frac{1}{2}$. This is a contradiction.

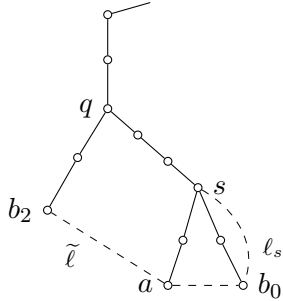

FIGURE 11. Illustration of $\widetilde{\ell}$ and $\ell_s$ in the proof of Lemma 8.6.

Now, we may assume that there exists a buddy link $\widetilde{\ell} \in \bigcup_{b \in L^{bud}(s)} \{buddylk(b)\}$ such that $x(\widetilde{\ell}) > 0$. W.l.o.g., let the original ends of $\widetilde{\ell}$ be $a$ and $b_2$. Let $q$ be the least common ancestor of $a$ and $b_2$ w.r.t. $T$; see Figure 11 and also see Figure 2 in Section 2. Observe that $T'_v$ contains $q$ (either as an original node or within a compound node), because $T'_v$ has at least 3 leaves, whereas every proper subtree of $T_q$ has $\leq 2$ leaves. Thus, $T'_v$ contains both ends of $\widetilde{\ell}$. Note that $b_2$ (the end of $\widetilde{\ell}$ not in $T'_s$) may be an original leaf of $T'_v$ or it may be contained in a compound leaf of $T'_v$ (since $C(T'_v) = \emptyset$, $b_2$ cannot be contained in a non-leaf compound node of $T'_v$).

Observe that $\ell_s \notin \delta_E^{\text{UP}}(s)$, otherwise, $\ell_s, \widetilde{\ell}$ form an overlapping pair (since $x(\widetilde{\ell}) > 0, x(\ell_s) = 1$). By similar arguments, $\ell_s$ cannot have both ends in $P'_{a,s}$. Thus, $\ell_s = sb_0$ ($\ell_s$ has an end at an $M$-exposed leaf of $T'_s$ because $\ell_s \in \delta_E^{\text{DN}}(s)$ and $\ell_s \in AM(T'_v)$). Let $\bar{\ell}_q$ denote the link in $AM(T'_v)$ that covers the tree-edge between $q$ and its parent (possibly, $q = v$, and in this case, $\bar{\ell}_q$ is the link incident with $\bar{v}$ in $AM(T'_v)$). Note that $\bar{\ell}_q$ has an end at a leaf in $T'_q$. Then, $\bar{\ell}_q$ forms an overlapping pair with either $\ell_s$ or $\widetilde{\ell}$ (since $x(\bar{\ell}_q) = 1$, $x(\ell_s) = 1$, $x(\widetilde{\ell}) > 0$). This contradicts the overlapping constraints of $(LP_0)$. This completes the proof of the lemma.

□

The following theorem is the key result of this paper. It is used only once; see the last paragraph of the proof of Theorem 8.8. Note that the last alternative in Theorem 8.7 (item (3)) is addressed in the last paragraph of the proof of Theorem 8.8.

**Theorem 8.7.** *Suppose that no (up-to-5) greedy contractions are applicable. Let $T'_v$ be a semiclosed tree that is not good. Then one of the following holds for $T'_v$.*

(1) $T'_v$ *is a deficient 3-leaf tree.*
(2) $T'_v$ *is a deficient 4-leaf tree.*
(3) $T'_v$ *has 4 leaves with $|M(T'_v)| = 1$, and moreover $T'_v$ has no cover of size 3.*

*Proof.* Since $T'_v$ is not good, Lemma 7.3 and Lemma 8.6(1) imply that $T'_v \neq T'$, $C(T'_v) = \emptyset$, and $|M(T'_v)| = 1$.

Observe that $T'_v$ has at least one $M$-exposed leaf. Otherwise, since $|M(T'_v)| = 1$, $T'_v$ has exactly two leaves and there exists an $M$-link between these two leaves; moreover, by Lemma 7.2(1), every node on the path of $T'$ between these two leaves is original; it follows that the link in $M(T'_v)$ is a twin link; this contradicts the definition of $M$. Thus, $T'_v$ has an $M$-exposed leaf and exactly two $M$-covered leaves. Hence, $T'_v$ has at least three leaves.

Also, observe that $|\mathcal{S} \cap V(T'_v)| \leq 2$. Otherwise, suppose that $|\mathcal{S} \cap V(T'_v)| \geq 3$. Then, by Lemma 8.2(1), for every stem $s \in \mathcal{S} \cap V(T'_v)$, every node in $T'_s$ is original; hence, there exists a stem $s^* \in \mathcal{S} \cap V(T'_v)$ such that both the leaves of $T'_{s^*}$ are $M$-exposed and there exists a twin link between these two leaves; this contradicts Lemma 7.2(2).

Let $\hat{e}$ denote the tree-edge between $v$ and its parent. Let $J$ denote $\{\ell \in \delta_E(u) \,:\, u \in (\mathcal{S} \cap V(T'_v)) \cup L^{matched}(T'_v) \cup \mathcal{B}(T'_v)\}$, where $\mathcal{B}(T'_v) = \mathcal{B}^{orig}(T'_v) \cup \mathcal{B}^{comp}(T'_v)$ (see the discussion before Lemma 8.4 for the definitions of $\mathcal{B}^{orig}(T'_v)$ and $\mathcal{B}^{comp}(T'_v)$).

By Lemma 8.4, $y$ can be written as a convex combination $\sum_{i \in Z} \lambda_i x^i$ such that $x^i \in \text{Las}_3(LP_0)$ and $x^i|_J$ is integral. Since $T'_v$ is not good and $|M(T'_v)| = 1$, Lemma 7.3 implies that $\Phi(y, T'_v) < \frac{1}{2}$. Thus, there exists an $i_0 \in Z$ such that $\Phi(x^{i_0}, T'_v) < \frac{1}{2}$.

We claim that $|U(T'_v)| \leq |\mathcal{S} \cap V(T'_v)| + 1$. To see this, suppose that $|U(T'_v)| \geq |\mathcal{S} \cap V(T'_v)| + 1$. Then, all of the conditions of Lemma 8.6(2) apply, hence, the lemma implies that $|U(T'_v)| = |\mathcal{S} \cap V(T'_v)| + 1$. Our claim follows.

Now, we analyze a few cases, depending on the number of leaves of $T'_v$. By Lemma 8.2(1), for every stem $s \in \mathcal{S} \cap V(T'_v)$, every node in $T'_s$ is original; thus, each stem $s \in \mathcal{S} \cap V(T'_v)$ contributes two original leaves to $T'_v$. Therefore, $|\mathcal{S} \cap V(T'_v)| \leq |L(T'_v)|/2$.

**Case 1:** $T'_v$ has exactly three leaves. Let the three leaves be $a, b_1, b_2$, where $a$ is $M$-exposed and $b_1, b_2$ are $M$-covered; thus, $b_1 b_2$ is the unique link in $M(T'_v)$, and $b_1, b_2$ are original nodes by Lemma 7.2(1). Clearly, $|\mathcal{S} \cap V(T'_v)| \leq 1$. We have two subcases, depending on $|\mathcal{S} \cap V(T'_v)|$.

**Subcase 1.1:** $|\mathcal{S} \cap V(T'_v)| = 1$. Let $s$ be the stem in $T'_v$. The $M$-link $b_1 b_2$ cannot have both ends in $T'_s$ (since $M$ contains no twin links). Thus, the $M$-exposed leaf $a$ is a leaf of $T'_s$. Then, it can be seen that $a$ is a bud and $buddylk(a) = b_1 b_2$ (see the definition of bud in Section 2). This is a contradiction, since $M$ contains no buddy links.

**Subcase 1.2:** $|\mathcal{S} \cap V(T'_v)| = 0$. Then we have $|U(T'_v)| = 1 = 1 + |\mathcal{S} \cap V(T'_v)|$. By Lemma 8.6(2)(i–iv), there exist two links $\ell_v \in \delta_E(\hat{e})$ and $\ell_a \in \delta_E(a)$ such that $x^{i_0}(\ell_v) = x^{i_0}(\ell_a) = 1$, these two links cover $T'_v$, and moreover, each of $b_1, b_2$ is incident to exactly one of these two links (since the auxiliary graph has a perfect matching formed by these two links).

If there is only one non-leaf node with degree other than 2 in $T'_v$ (see Figure 6(a)), then $T'_v$ is a deficient 3-leaf tree. We are done. Otherwise, we have exactly two non-leaf nodes $u, q$ in $T'_v$ with degree other than 2. In fact, both these nodes have degree 3 since $T'_v$ has exactly 3 leaves. W.l.o.g., let $u$ be a proper ancestor of $q$. Then, $T'_q$ has only two leaves. By the argument at the beginning of the proof, the $M$-link in $T'_v$ cannot connect both leaves in $T'_q$. This implies that one leaf of $T'_q$ is $M$-exposed. Thus, $a$ is a leaf of $T'_q$. W.l.o.g., let the other leaf of $T'_q$ be $b_2$. Then, $b_1$ is the third leaf, which is not in $T'_q$.

Suppose that $\ell_v$ is incident to $b_1$ and $\ell_a$ is incident to $b_2$. Then, the tree-edge between $q$ and its parent is not covered by these two links (see Figure 12(a)). This is a contradiction. Hence, $\ell_v$ is incident to $b_2$ and $\ell_a$ is incident to $b_1$ (see Figure 12(b)). Therefore, $T'_v$ satisfies all the conditions of a deficient 3-leaf tree.

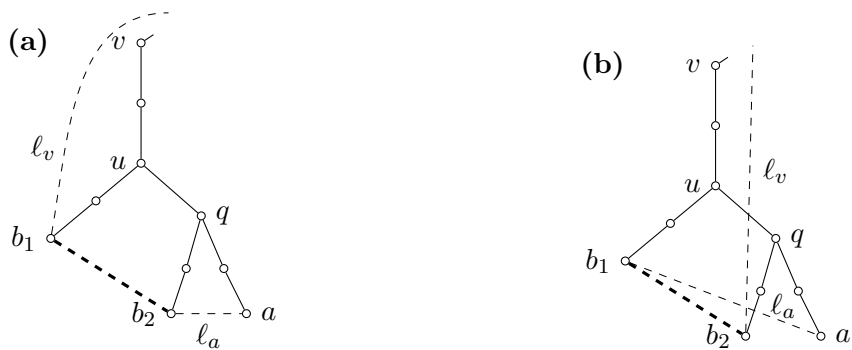

FIGURE 12. The links $\ell_v$ and $\ell_a$ in Subcase 1.2 of the proof of Theorem 8.7. The dashed lines indicate links and the thick dashed lines indicate $M$-links.

**Case 2:** $T'_v$ has exactly four leaves. Let the four leaves be $a_1, a_2, b_1, b_2$, where $a_1, a_2$ are $M$-exposed and $b_1, b_2$ are $M$-covered; thus, $b_1 b_2$ is the unique link in $M(T'_v)$, and $b_1, b_2$ are original nodes by Lemma 7.2(1). We have $1 = |U(T'_v)| - 1 \leq |\mathcal{S} \cap V(T'_v)| \leq 2$. As above, we have two subcases, depending on $|\mathcal{S} \cap V(T'_v)|$.

**Subcase 2.1:** $|\mathcal{S} \cap V(T'_v)| = 1$. Let $s$ be the stem in $T'_v$. By Lemma 7.2(2), it can be seen that the twin link of $s$ is incident to one $M$-exposed leaf, say $a_1$, and to one $M$-covered leaf, say $b_1$ (see Subcase 1.1).

Note that $|U(T'_v)| = 2 = 1 + |\mathcal{S} \cap V(T'_v)|$. By Lemma 8.6(2)(i–iv), there exist three links $a_1 b_1$ (the twin link of $s$), $\ell_v \in \delta_E(\hat{e})$ and $\ell_{a_2} \in \delta_E(a_2)$ such that $x^{i_0}(a_1 b_1) = x^{i_0}(\ell_v) = x^{i_0}(\ell_{a_2}) = 1$, these three links cover $T'_v$, and moreover, each of $s, b_2$ is incident to exactly one of the two links $\ell_v, \ell_{a_2}$ (since the auxiliary graph has a perfect matching formed by the three links); moreover, $s$ cannot be incident to $\ell_v$ (by Lemma 8.6(2)(iv)). Thus $\ell_v$ is incident to $b_2$, and $\ell_{a_2}$ is incident to $s$, i.e., $\ell_{a_2} = a_2 s$.

Let $p$ be the least common ancestor of $s$ and $a_2$ (in $T'_v$). If $T'_p$ does not contain all the leaves of $T'_v$, then it can be seen that the tree-edge between $p$ and its parent is not covered by the three links $a_1 b_1, a_2 s, \ell_v$ (see Figure 13(a)). This is a contradiction (since $T'_v$ is covered by $\{a_1 b_1, a_2 s, \ell_v\}$). It follows that $T'_p$ contains all the leaves of $T'_v$. Then, it can be seen that $T'_v$ is a deficient 4-leaf tree (see Figure 13(b)).

**Subcase 2.2:** $|\mathcal{S} \cap V(T'_v)| = 2$. If $T'_v$ has no cover of size 3, then item (3) in the statement of the theorem holds. Thus, we may assume that $T'_v$ has a cover of size 3.

Let $s_1, s_2$ be the two stems in $T'_v$. Let $a_1, b_1$ be the original leaves of $T'_{s_1}$, and let $a_2, b_2$ be the original leaves of $T'_{s_2}$. We may assume that $a_1, a_2$ are $M$-exposed and $b_1, b_2$ are $M$-covered, because each twin link is incident to one $M$-exposed leaf and to one

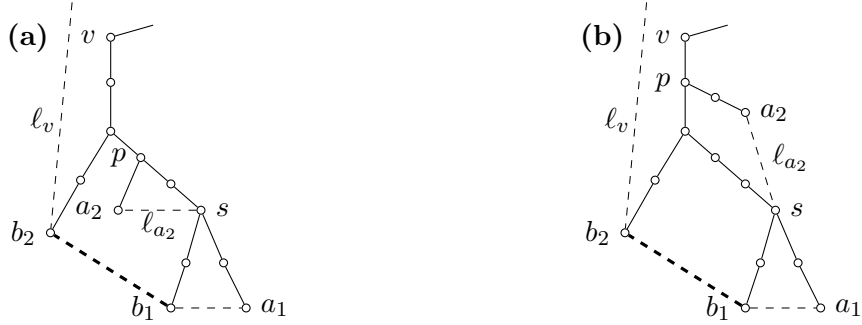

FIGURE 13. The links $\ell_v$, $\ell_{a_2}$ and $a_1b_1$ in subcase 2.1 of the proof of Theorem 8.7. The dashed lines indicate links and the thick dashed lines indicate $M$-links. Another case could occur, but it is not shown here (see Figure 7).

$M$-covered leaf by Lemma 7.2(2) (see Subcase 2.1). Thus, $b_1b_2$ is the unique link in $M(T'_v)$.

By Lemma 8.2(1) and $C(T'_v) = \emptyset$, $T'_v$ has no compound nodes, i.e., all its nodes are original.

Observe that $T'_v$ has at least $3\frac{1}{2}$ credits from the two $M$-exposed leaves and the $M$-link.

Consider the possible leaf-to-leaf links of $T'_v$. The link $a_1a_2$ does not exist, by Lemma 7.2(2), since $a_1, a_2$ are both $M$-exposed. If both the links $a_1b_2$ and $a_2b_1$ exist, then an (up-to-5) greedy contraction applies (we contract these two links and we have $3\frac{1}{2}$ credits). Thus, at most one of the links $a_1b_2$ or $a_2b_1$ is present.

If $T_v$ has no leafy 3-cover w.r.t. $T$, then it satisfies all the conditions for a bad 2-stem tree, hence, it would have been contracted in Preprocessing step 1 ($\Lambda$-contraction). This is a contradiction.

Thus, we may assume that $T_v$ has a leafy 3-cover $\widetilde{J} = \{\ell_0, \ell_1, \ell_2\}$ w.r.t. $T$, where $\ell_0 = uw$ is a link such that one end $u$ is in $T_v$ and the other end $w$ is a leaf in $L - L(T_v)$. Let $\ell'_0 = uw'$ be the link w.r.t. $T'$ that corresponds to $\ell_0$, where $w' = w$ if $w$ is still an original node in $T'$, and otherwise, $w'$ is the compound node containing $w$. Note that in either case, $w'$ is not in $V(T'_v)$ since every node in $T'_v$ is original. If $w'$ is a compound node or an $M$-exposed original leaf, then an (up-to-5) greedy contraction applies (we contract the 3 links in $\widetilde{J}$ and we have $4\frac{1}{2}$ credits). Otherwise, if $w'$ is an $M$-covered original leaf, then again an (up-to-5) greedy contraction applies (we contract 4 links, namely, the links in $\widetilde{J}$ and the $M$-link incident to $w$, and we have $3\frac{1}{2} + \frac{3}{2} = 5$ credits). Thus, the existence of a leafy 3-cover gives a contradiction.

**Case 3:** $T'_v$ has at least five leaves. Then, $T'_v$ has at least three $M$-exposed leaves and two $M$-covered leaves. Thus, we have $3 \leq |U(T'_v)| \leq |\mathcal{S} \cap V(T'_v)| + 1 \leq 3$, since $|\mathcal{S} \cap V(T'_v)| \leq 2$; it follows that $|U(T'_v)| = 3$, $|\mathcal{S} \cap V(T'_v)| = 2$, and $T'_v$ has exactly 5 leaves.

Let $s_1, s_2$ be the two stems in $T'_v$. Let $a_1, b_1$ be the original leaves of $T'_{s_1}$, and let $a_2, b_2$ be the original leaves of $T'_{s_2}$. We may assume that $a_1, a_2$ are $M$-exposed and $b_1, b_2$ are $M$-covered, because each twin link is incident to one $M$-exposed leaf and to one $M$-covered leaf by Lemma 7.2(2) (see Subcase 2.1). Thus, $b_1b_2$ is the unique link in $M(T'_v)$. Let $u$ be the fifth leaf of $T'_v$, where $u$ is not in $T'_{s_1}$ nor in $T'_{s_2}$.

Since $|U(T'_v)| = 3 = 1 + |\mathcal{S} \cap V(T'_v)|$, Lemma 8.6(2)(i–iv) implies that there exist four links $\ell_v \in \delta_E(\hat{e})$, $\ell_u \in \delta_E(u)$, $a_1b_1$, $a_2b_2$ (the twin links of $s_1, s_2$) such that each of $s_1, s_2$ is incident to exactly one of the two links $\ell_v, \ell_u$ (since the auxiliary graph has a perfect matching formed by these four links); moreover, $\ell_v$ cannot be incident to $\{s_1, s_2\}$ (by Lemma 8.6(2)(iv)). This is impossible. Thus, $T'_v$ cannot have more than four leaves.

The theorem follows from the above case analysis. □

## 8.3. Addressing deficient trees.

This section discusses the missing piece of the algorithm, namely, the handling of deficient trees. Even, et al., see [5, Section 4.3], presented an elegant method for addressing deficient 3-leaf trees. We use essentially the same method in Part I (Section 7.2). The key point is to compute another (auxiliary) matching of the leaf-to-leaf links of $T'$, and then to find a minimally semiclosed tree $T'_v$ w.r.t. this auxiliary matching; moreover, we prove (in Theorem 7.4 of Part I) that $T'_v$ is a good semiclosed tree w.r.t. the "original matching," and $T'_v$ has a fitting cover of appropriate size. In this section, we extend this method to address deficient 3-leaf trees as well as deficient 4-leaf trees. The extension is straightforward (but non-trivial). We start by "transforming" the current tree $T'$ to another tree $\widetilde{T}$; this is achieved by applying "latch contractions" to the latches of deficient 4-leaf trees in an appropriate way (the details are discussed below). Then, we apply the method from Part I (Section 7.2) to $\widetilde{T}$. Thus, the main goal in this section is to find a semiclosed tree $T'_v$ w.r.t. $M$ of the current tree $T'$ such that $T'_v$ is good (i.e., it has $\geq |\Gamma(M, T'_v)| + 1$ credits) and $T'_v$ has a fitting cover of size $|\Gamma(M, T'_v)|$.

For a deficient 3-leaf tree $T'_v$, if $T'_v$ is not a proper subtree of another deficient 3-leaf tree, then we call $T'_v$ a *maximal deficient* 3-*leaf tree*. Similarly, we define a *maximal deficient* 4-*leaf tree*. By Property 2.1, any two different maximal deficient 3-leaf trees are disjoint; similarly, any two different maximal deficient 4-leaf trees are disjoint.

Let $E^{\text{latch}}$ denote the set of latches of all maximal deficient 4-leaf trees of $T'$. Let $\widetilde{T} = T'/E^{\text{latch}}$. We mention that $\widetilde{T}$ is an auxiliary graph that is used for achieving the main goal of this section (find a semiclosed tree $T'_v$ with desired properties), and other than that, the algorithm does not refer to $\widetilde{T}$ at all; thus, the algorithm does not add any latch to $F$ while constructing $\widetilde{T}$, nor does it incur any credits while constructing $\widetilde{T}$. By a *latched-node* we mean a node of $\widetilde{T}$ that results from the contraction of a latch (note that these are not compound nodes of $T'$, and the algorithm does not assign any credits to a latched-node).

Since two different maximal deficient 4-leaf trees are disjoint, each "latch contraction" results in a distinct latched-node; thus, there is a bijection between the latches of $T'$ and the latched-nodes of $\widetilde{T}$. For a subtree $\widetilde{T}_w$ of $\widetilde{T}$, we use $\text{latch}(\widetilde{T}_w)$ to denote the set consisting of latches that correspond to latched-nodes of $\widetilde{T}_w$.

For a node $w$ of $T'$, let $\widetilde{w}$ denote the node of $\widetilde{T}$ that corresponds to $w$. If there exists no latch $cs$ of $T'$ such that $w$ is in $P'_{c,s}$ (the path of $T'$ between $c, s$), then $\widetilde{w} = w$, otherwise, $\widetilde{w}$ is the latched-node formed by contracting the (unique) latch $cs$ such that $w$ is in $P'_{c,s}$ (in this case, the latch $cs$ is unique because two different maximal deficient 4-leaf trees are disjoint).

Then, as in Part I (Section 7.2), we construct an auxiliary matching of the leaf-to-leaf links of $\widetilde{T}$ denoted by $M^{\text{new}}$. For ease of exposition, we denote the image (in $\widetilde{T}$) of $M(T')$ by $M(\widetilde{T})$, and we refer to the image (in $\widetilde{T}$) of a link in $M(T')$ as an $M$-link w.r.t. $\widetilde{T}$ (this is justified because any $M$-link w.r.t. $T'$ has no ends in common with the nodes of $T'$ that get contracted into latched-nodes). To construct $M^{\text{new}}$, we start with $M^{\text{new}} := M$, then we examine each maximal deficient 3-leaf tree $\widetilde{T}_{\widetilde{w}}$ of $\widetilde{T}$ and we replace the unique link of $M(\widetilde{T}_{\widetilde{w}})$ in $M^{\text{new}}$ by another leaf-to-leaf link. In more detail, consider any maximal deficient 3-leaf tree $\widetilde{T}_{\widetilde{w}}$, and let the three leaves be $a, b, d$, where $a$ is $M$-exposed, $b$ is the ceiling leaf, and $bd$ is the unique link in $M(\widetilde{T}_{\widetilde{w}})$; we remove the $M$-link $bd$ from $M^{\text{new}}$ and instead we add the link $ad$ to $M^{\text{new}}$ (see Algorithm 2). Since any two different maximal deficient 3-leaf trees are disjoint, this replacement takes place independently for each maximal deficient 3-leaf tree. We mention that $M^{\text{new}}$ is an auxiliary matching that is used for achieving the main goal of this section (find a semiclosed tree $T'_v$ with desired properties), and other than that, the algorithm does not refer to $M^{\text{new}}$ at all; whereas, the matching $M$ (and its image $M(T')$) are used throughout the algorithm and its analysis.

**Theorem 8.8.** *Suppose that no (up-to-5) greedy contractions are applicable. Let $\widetilde{T} = T'/E^{\text{latch}}$, and let $\widetilde{T}_{\widetilde{v}}$ be a minimally semiclosed tree w.r.t. $M^{\text{new}}$. Then, $\widetilde{v}$ cannot be a latched-node of $\widetilde{T}$; thus, $\widetilde{v} = v$, where $v$ is a node of $T'$. Moreover, $T'_v$ is a good semiclosed tree w.r.t. $M$ and $T'_v$ has a fitting cover $\Gamma(M^{\text{new}}, \widetilde{T}_v) \cup \text{latch}(\widetilde{T}_v)$ of size $|\Gamma(M, T'_v)|$.*

*Proof.* We start by addressing the first statement. By way of contradiction, suppose that $\widetilde{v}$ is a latched-node. Observe that every latched-node is contained in a deficient 3-leaf tree of $\widetilde{T}$ (a latched-node is formed by a "latch contraction" that results in a deficient 3-leaf tree). Hence, $\widetilde{T}_{\widetilde{v}}$ is a subtree of a maximal deficient 3-leaf tree of $\widetilde{T}$. But, by our construction of $M^{\text{new}}$, the ceiling leaf of every maximal deficient 3-leaf tree is $M^{\text{new}}$-exposed, hence, no subtree of a maximal deficient 3-leaf tree is semiclosed w.r.t. $M^{\text{new}}$. Thus, $\widetilde{T}_{\widetilde{v}}$ is not semiclosed w.r.t. $M^{\text{new}}$. This is a contradiction.

It follows that $\widetilde{v}$ cannot be a latched-node. Thus, there exists a node $v$ in $T'$ such that $v = \widetilde{v}$; moreover, $\widetilde{T}_{\widetilde{v}} = \widetilde{T}_v$, and $T'_v$ is well defined.

**Claim 8.9.** *(1) If $T'_v$ has a node in a maximal deficient 4-leaf tree of $T'$, then $T'_v$ properly contains this maximal deficient 4-leaf tree.*
*(2) If $T'_v$ has a node in a maximal deficient 3-leaf tree of $T'$, then $T'_v$ properly contains this maximal deficient 3-leaf tree.*

We address statement (1) of the claim. Suppose that $T'_v$ has a node in a maximal deficient 4-leaf tree $T'_w$. By "latch contraction," $T'_w$ is transformed to a deficient 3-leaf tree $\widetilde{T}_{\widetilde{w}}$ of $\widetilde{T}$. Thus, $\widetilde{T}_v$ has a node in a maximal deficient 3-leaf tree $\widetilde{T}_{\widetilde{u}}$, where $\widetilde{u}$ is an ancestor of $\widetilde{w}$ in $\widetilde{T}$. But, (as discussed above) no subtree of $\widetilde{T}_{\widetilde{u}}$ is semiclosed w.r.t. $M^{\text{new}}$. Hence, $\widetilde{T}_v$ cannot be a subtree of $\widetilde{T}_{\widetilde{u}}$. Then Property 2.1 implies that $\widetilde{T}_v$ properly contains $\widetilde{T}_{\widetilde{u}}$. Hence, $v$ is a proper ancestor of $w$ in $T'$, and $T'_v$ properly contains $T'_w$.

Next, we address statement (2) of the claim. Suppose that $T'_v$ has a node in a maximal deficient 3-leaf tree $T'_w$. If this maximal deficient 3-leaf tree is contained in a maximal deficient 4-leaf tree, then we are done by the previous discussion. Hence, by Property 2.1, we may assume that $T'_w$ has no node in a maximal deficient 4-leaf tree. Then $\widetilde{T}_{\widetilde{w}}$ contains no latched-node, and $\widetilde{T}_{\widetilde{w}} = \widetilde{T}_w = T'_w$ is a maximal deficient 3-leaf tree of $\widetilde{T}$. Since $\widetilde{T}_v$ has a node in $\widetilde{T}_w$, and no subtree of $\widetilde{T}_w$ is semiclosed w.r.t. $M^{\text{new}}$, Property 2.1 implies that $\widetilde{T}_v$ properly contains $\widetilde{T}_w$. It follows that $T'_v$ properly contains $T'_w$. This proves the claim.

Claim 8.9 implies that $T'_v$ is neither a deficient 3-leaf tree nor a deficient 4-leaf tree.

Next, we address the second statement in the theorem. To show that $T'_v$ is semiclosed w.r.t. $M$, we must show that every link incident to an $M$-exposed leaf of $T'_v$ has both ends in $T'_v$, and every $M$-link incident to $T'_v$ has both ends in $T'_v$. Let $u$ be an $M$-exposed leaf of $T'_v$. Suppose that $u$ is contained in $T'_w$, where $T'_w$ is either a deficient 3-leaf tree or a deficient 4-leaf tree (w.r.t. $T'$). Then $T'_v$ contains $T'_w$, by Claim 8.9. Since $T'_w$ is a semiclosed tree w.r.t. $M$ (by definition of deficient trees), all links incident with $u$ have both ends in $T'_w$, hence, also in $T'_v$. Next, suppose that $u$ does not belong to any deficient 3-leaf tree or any deficient 4-leaf tree (w.r.t. $T'$). Such an $M$-exposed node of $T'_v$ is an $M^{\text{new}}$-exposed node of $\widetilde{T}$. Since $\widetilde{T}_v$ is semiclosed w.r.t. $M^{\text{new}}$, it follows that $\widetilde{T}_v$ contains both ends of each link incident with $\widetilde{u} = u$; hence, $T'_v$ contains both ends of each link incident with $u$. Next, consider an $M$-link $uw$ incident to $T'_v$, where $u$ is a node of $T'_v$. If $u$ is contained in $T'_w$, where $T'_w$ is a deficient 3-leaf tree or a deficient 4-leaf tree, then note that $T'_w$ is semiclosed, and moreover, $T'_w$ is properly contained in $T'_v$, hence, both ends of $uw$ are in $T'_v$. Otherwise, $uw$ has no end in a deficient 3-leaf tree or in a deficient 4-leaf tree; then, by our construction of $M^{\text{new}}$, we have $uw \in M^{\text{new}}$ ($M$ and $M^{\text{new}}$ are the same, except "inside" some deficient tree). Since $\widetilde{T}_v$ is semiclosed w.r.t. $M^{\text{new}}$, it follows that $uw$ has both ends in $\widetilde{T}_v$, hence, $uw$ has both ends in $T'_v$. Thus, $T'_v$ is semiclosed w.r.t. $M$.

By the construction of $\widetilde{T}$ and $M^{\text{new}}$, Claim 8.9, and the fact that $\widetilde{T}_v$ is semiclosed, we have $|M(T'_v)| = |M^{\text{new}}(\widetilde{T}_v)|$; moreover, the number of $M$-exposed nodes of $T'_v$ is equal to the sum of the number of $M^{\text{new}}$-exposed nodes of $\widetilde{T}_v$ and $|latch(\widetilde{T}_v)|$ (since each latched-node of $\widetilde{T}_v$ corresponds to an $M$-exposed leaf of a maximal deficient 4-leaf tree of $T'_v$). Hence, we have $|\Gamma(M, T'_v)| = |\Gamma(M^{\text{new}}, \widetilde{T}_v)| + |latch(\widetilde{T}_v)|$. Clearly, $\Gamma(M^{\text{new}}, \widetilde{T}_v)$ is a fitting cover of $\widetilde{T}_v$, because $\widetilde{T}_v$ is a minimally semiclosed tree w.r.t. $M^{\text{new}}$ (by Lemma 6.2). It follows that $\Gamma(M^{\text{new}}, \widetilde{T}_v) \cup latch(\widetilde{T}_v)$ is a fitting cover of $T'_v$ of size $|\Gamma(M, T'_v)|$.

Finally, we apply Theorem 8.7 to prove that $T'_v$ is good. Suppose that $T'_v$ is not good. By Claim 8.9, $T'_v$ is neither a deficient 3-leaf tree nor a deficient 4-leaf tree. Then, Theorem 8.7(3) implies that $T'_v$ has 4 leaves, $|M(T'_v)| = 1$, and $T'_v$ has no cover of size 3. Since $|M(T'_v)| = 1$ and $|U(T'_v)| = 2$, we have $|\Gamma(M, T'_v)| = 3$. The previous paragraph shows that $T'_v$ has a fitting cover of size $|\Gamma(M, T'_v)| = 3$. Thus, we have a contradiction. It follows that $T'_v$ is good. This completes the proof. □

The procedure described above to find a good semiclosed tree is summarized in the following pseudocode.

---

let $E^{latch}$ denote the set of latches of all maximal deficient 4-leaf trees of $T'$;
let $\widetilde{T} := T'/E^{latch}$;
start with $M^{\text{new}} := M$;
**for** *each maximal deficient 3-leaf tree $\widetilde{T}_{\widetilde{w}}$ in $\widetilde{T}$* **do**
  let $b$ be the ceiling leaf, $a$ be the $M$-exposed leaf and $db$ be the $M$-link in $\widetilde{T}_{\widetilde{w}}$;
  update $M^{\text{new}}$ by replacing $db$ by $da$ ($M^{\text{new}} := M^{\text{new}} - \{db\} \cup \{da\}$);
**end**
find a minimally semiclosed tree $\widetilde{T}_v$ w.r.t. $M^{\text{new}}$;
$T'_v$ is a good semiclosed tree w.r.t. $M$ with a fitting cover $\Gamma(M^{\text{new}}, \widetilde{T}_v) \cup latch(\widetilde{T}_v)$ of size $|\Gamma(M, T'_v)|$ (by Theorem 8.8);
return $T'_v$ and its fitting cover $\Gamma(M^{\text{new}}, \widetilde{T}_v) \cup latch(\widetilde{T}_v)$;
  **Algorithm 2:** Find a good semiclosed tree by addressing deficient trees.

---

Algorithm 2 shows how to find a good semiclosed tree $T'_v$ together with a fitting cover of size $|\Gamma(M, T'_v)|$ for the main loop in Algorithm 1 in Section 7. (The first step of Algorithm 2 applies "latch contractions" to construct the auxiliary tree $\widetilde{T}$ from $T'$; "latch contractions" have no effect on $F$ or on the credits of the algorithm.)

In conclusion, Algorithm 1 runs in polynomial time and returns a solution for TAP of size $\leq (\frac{3}{2} + \epsilon)y(E)$. This proves Theorem 6.1.

Although our integrality ratio (and approximation guarantee) is proved relative to a "weaker lower bound" than the optimum value (since the feasible region of a relaxation is a superset of the convex full of integer solutions), we loose an additive term of $\epsilon$ in the approximation guarantee, compared to the $\frac{3}{2}$ approximation guarantee of [10]. But, we can show that our methods achieve an approximation guarantee of $\frac{3}{2}$ relative to the (integral) optimum value. Observe that the algorithm is combinatorial and does not refer to the SDP relaxation at all (our analysis does rely on the SDP relaxation and properties of its solutions). Thus, we can apply the algorithm and analysis for the level $t = |E|$ tightening of $(LP_0)$, and at this level, the set of (feasible) solutions of $\text{Las}_t(LP_0)$ corresponds to the convex hull of integer solutions. Consequently, for level $t = |E|$, Lemma 4.4 can be restated as follows: if $y \in \text{Las}_t(LP_0)$, where $t = |E|$, then $y|_{E(L)}$ is in the matching polytope of $G(L) = (L, E(L))$. Hence, for level $t = |E|$, the upperbound on the potential function stated in Lemma 5.2 can be improved from $(\frac{3}{2} + \epsilon)y(E)$ to $\frac{3}{2}y(E)$. In other words, for $y \in \text{Las}_t(LP_0)$, where $t = |E|$, the potential function is $\leq \frac{3}{2}y(E)$. Sections 6–8 show that Algorithm 1 returns a solution for TAP whose size is $\leq$ the potential function. Thus, we have the following result.

**Corollary 8.10.** *Let $y$ denote an optimal (integral) solution of TAP. Algorithm 1 runs in polynomial time, and returns a solution for TAP of size $\leq \frac{3}{2}y(E)$.*

It can be seen that Algorithm 1 can be implemented to run in time $O(|V|^3 + |V|^2 m_0^5)$, where $m_0$ denotes the number of (multi) links in the input and we assume that $m_0 = O(|V|^2)$. This can be achieved using simple data structures such as adjacency matrices and adjacency lists; the details are not straightforward, but we skip them. To handle the (up-to-5) greedy contractions, every subset of size $\leq 5$ of the set of maximal links has to be examined (for each link $uw$ in a subset, $P_{u,w}$ has to be examined), and there are $O(|V|)$ such contractions, hence, these contractions contribute a total of $O(|V|^2 m_0^5)$ to the running time.

**Theorem 8.11.** *Algorithm 1 can be implemented to run in time $O(|V|^3 + |V|^2 m_0^5)$.*

**Acknowledgments.** We thank André Linhares for many discussions. We thank several other colleagues who read preliminary drafts and gave us insightful comments.

Dept. of Comb. & Opt., University of Waterloo, Waterloo, Ontario N2L3G1, Canada.
*E-mail address*: `jcheriyan@uwaterloo.ca`

*E-mail address*: `z9gao@uwaterloo.ca`